# Mixed Quantum/Classical Theory (MQCT) Approach to the Dynamics of Molecule-Molecule Collisions in Complex Systems

Carolin Joy, Bikramaditya Mandal, Dulat Bostan, Marie-Lise Dubernet[1] and Dmitri Babikov[*]

**Abstract:** We developed a general theoretical approach and a user-ready computer code that permit to study the dynamics of collisional energy transfer and ro-vibrational energy exchange in complex molecule-molecule collisions. The method is a mixture of classical and quantum mechanics. The internal ro-vibrational motion of collision partners is treated quantum mechanically using time-dependent Schrodinger equation that captures many quantum phenomena including state quantization and zero-point energy, propensity and selection rules for state-to-state transitions, quantum symmetry and interference phenomena. A significant numerical speed up is obtained by describing the translational motion of collision partners classically, using the Ehrenfest mean-field trajectory approach. Within this framework a family of approximate methods for collision dynamics is developed. Several benchmark studies for diatomic and triatomic molecules, such as $H_2O$ and $ND_3$ collided with He, $H_2$ and $D_2$, show that the results of MQCT are in good agreement with full-quantum calculations in a broad range of energies, especially at high collision energies where they become nearly identical to the full quantum results. Numerical efficiency of the method and massive parallelism of the MQCT code permit us to embrace some of the most complicated collisional systems ever studied, such as $C_6H_6$ + He, $CH_3COOH$ + He and $H_2O$ + $H_2O$. Application of MQCT to the collisions of chiral molecules such as $CH_3CHCH_2O$ + He, and to the molecule-surface collisions is also possible and will be pursued in the future.

[1] Observatoire de Paris, PSL University, Sorbonne Université, CNRS, LERMA, Paris, France
[*] Author to whom all correspondence should be addressed; electronic mail: dmitri.babikov@mu.edu
Chemistry Department, Wehr Chemistry Building, Marquette University, Milwaukee, Wisconsin 53201-1881, USA

## I. INTRODUCTION

The idea of simplifying the description of complex molecular systems by mixing classical and quantum theories in one hybrid method is not entirely new or unique. For example, the electronic structure calculations of complex molecular environments are greatly facilitated by the QM/MM method,[1,2] that employs classical force-field (MM) for the description of a large part of the molecule, while the *ab initio* calculations (QM) are carried out for a small subsystem (where an accurate description of chemical transformations is more important). Another example can be drawn from statistical mechanics, where the translational and rotational partition functions are computed in the high-temperature (classical) limit, while the vibrational and electronic wavefunctions appeal to the quantization of states.[3,4] Also, the calculations of chemical reaction rates by propagating classical trajectories on a global potential energy surface are carried out by neglecting quantum effects, but then a tunneling correction is applied *a posterior* [5–7] by treating quantum mechanically a small part of configuration space near the top of reaction barrier. Finally, the mean-field Ehrenfest approach to the modelling of electronically non-adiabatic processes is another directly relevant example.[8–11] More examples of successful combination of classical and quantum mechanics can be found in recent literature.[12–14]

It would be logical to assume that, in complex molecular systems, a similar strategy could be quite beneficial for the description of molecular dynamics in general, and for the calculations of inelastic molecular collisions in particular. It appears, however, that the quantum-classical methods for molecular dynamics were largely abandoned in early 2000s,[15–20] in favor of full-quantum and purely-classical methods. This happened, perhaps, due to an explosive development of computer clusters, which gave us hope that a brute force computer power may permit us to win the battle against quantum complexity. This did not happen. The development of full quantum calculations progresses, but they remain computationally challenging.[21–26] Classical trajectory calculations, on the other side, are computationally affordable but their predictive power is limited to state-averaged observables and is not free of several deficiencies.[27,28]

During the last decade we undertaken the development of a mixed quantum/classical theory (MQCT) for collisional energy transfer and ro-vibrational energy exchange between two molecular collision partners.[29–31] In this approach the translational motion of two molecules (their scattering process) is described classically using the approximate mean-field trajectory method (Ehrenfest), which gives a considerable computational speed up.[32,33] Quantum mechanics is employed for the description of their rotational and vibrational states, within the time-dependent Schrodinger equation formalism, which permits us to retain in the model the quantization of the internal states of collision partners with their intrinsic symmetry properties, selection rules, propensities to certain types of state-to-state transitions and zero-point energy conservation. We found that MQCT approach is both numerically affordable and physically accurate. In particular, we carried out MQCT calculations for a number of simple systems, such as diatom + atom,[34,35] triatomic + atom[36] and diatom + diatom,[37,38] where the full quantum calculations are still affordable and can be used as a solid benchmark. We showed that in these cases MQCT method gives reliable predictions of integral cross sections for various individual state-to-state transitions in a broad range of energies and for various collision partners, including the elastic scattering channel, but also permits to obtain differential cross sections in good agreement with full quantum results.[39] At high collision energies, where quantum calculations become challenging due to a large number of partial scattering waves, the MQCT calculations remain affordable and become quite accurate, often giving the results indistinguishable from those of the full quantum calculations. But even at low collision energies the results of MQCT remain reasonable, obeying threshold behavior for excitation of the individual quantum states and approximately satisfying the principle microscopic reversibility.[40,41] In addition, MQCT calculations give a valuable and unique time-dependent insight into the process, by illuminating time-evolution of state populations during the collision event.[41,42]

This recent successful work is very encouraging, still, one must remember that MQCT approach involves classical approximation for the translational motion of collision partners, and, like any approximate method, it has its limitations. In particular, at low collision energy, when the collision process is dominated by scattering resonances, MQCT trajectories become permanently trapped in the interaction region forming a collision complex (the analogue of Feshbach resonance) but the rigorous method for analysis of orbiting trajectories is yet to found.[39] Due to the same classical approximation, MQCT cannot describe the so-called “shape-type” resonances populated by quantum tunneling through the barrier. It is also unknown how to apply MQCT to reactive collisions, when some bonds are broken, and others are formed. Therefore, it is advised to test MQCT calculations against the full-quantum calculations, whenever it is possible, which is one of the goals of this paper.

We also tried to push MQCT calculations to the limit, by applying this method to more complex systems where the quantum calculations are next to impossible and the accurate benchmark data either do not exist at all or are very limited (e.g., available only for low collision energies and/or small rotational basis set, or only an approximate quantum treatment is affordable). Among these systems are medium size polyatomic molecules such as methyl formate $HCOOCH_3$, benzene $C_6H_6$ and polypropylene oxide $CH_3CHCH_2O$ collided with an atom,[40,42,43] and a collision of two general asymmetric-top-rotor molecules, such as $H_2O$ + $H_2O$ and their isotopic substitutions.[44,45] Some of these calculations are still ongoing but the results are encouraging both in terms of MQCT accuracy and numerical affordability (or practicality).

In this paper we push MQCT calculations to another limit, now in terms of the number of high-energy quantum states included in the model, and the range of collision energies covered, by applying this method to the rotationally inelastic scattering in $H_2O$ + $H_2$ system. This process is one of the critical energy transfer steps in the evolution of interstellar media (molecular clouds), star forming environments (proto-stellar accretion discs and hot pre-stellar cores) and stellar atmospheres at the late stage of start evolution (red-giants). In the past, accurate full quantum calculations were carried out for quenching of 44 rotationally excited levels of both para-$H_2O$ and ortho-$H_2O$ (90 levels of water total) combined with two states of $H_2$ projectile: $j = 0, 2, 4$ of para-hydrogen and $j = 1, 3$ of ortho-hydrogen (5 states total), at the center-of-mass collision energy up to 8000 cm$^{-1}$. It appears, however, that astrophysical modelers, the user of these data, want to significantly expand the range of temperature in their simulations, begging for broader range of collision energies and higher levels of rotational excitation of both collision partners.[46] In this work we present the results of MQCT using a rotational basis set that includes 100 states of para-$H_2O$ and ortho-$H_2O$ (200 states total) and all rotational states of $H_2$ projectile up to $j = 10$, for collision energies up to 12000 cm$^{-1}$, without invoking the coupled-states approximation,[46] i.e., retaining in the model the effect of Coriolis coupling.

There are several goals that we target here. In Sec. II we summarize the equations of motion used in MQCT and derive the formulae for state-to-state transition matrix in $H_2O$ + $H_2$ and any other similar system (asymmetric-top rotor + linear rotor). In Sec. III we give technical details of these calculations and conduct a comprehensive check of microscopic reversibility, using cross sections computed for the excitation and quenching directions of each individual state-to-state transition, for various collision energies. In Sec. IV we present a detailed comparison of the individual state-to-state transition rate coefficients computed by MQCT against the full-quantum results from literature,[47] through the temperature range available there. In Sec. V, we analyze the results of MQCT calculations for those highly excited states that were not covered previously, to obtain insight into the overall trends of state-to-state energy transfer processes in $H_2O$ + $H_2$ system including very high levels of rotational excitation and a broad range of collision energies. Conclusions are summarized in Sec. VI. More details are presented in Supplemental Information.

## II. THEORETICAL APPROACH

The rotations of each colliding partner are treated quantum mechanically and the wavefunction depends on the angles needed to describe individual orientations of these molecules. In general, for an asymmetric-top molecule the rotations are described by a set of Euler angles $\Lambda_1 = (\alpha_1, \beta_1, \gamma_1)$. Following Parker,[48] we use *active* rotations for each collision partner. The rotational states of the $H_2O$ molecule are quantized and are described by the following wavefunctions:

$$\psi_{j_1 m_1 \tau_1}(\alpha_1, \beta_1, \gamma_1) = \sqrt{\frac{2j_1 + 1}{8\pi^2}} \sum_{k=-j_1}^{+j_1} b_k^{\tau_1} D_{m_1,k}^{j_1 *}(\alpha_1, \beta_1, \gamma_1) \tag{1}$$

where the set of expansion coefficients $b_k^{\tau_1}$ is obtained by diagonalization of an asymmetric-top rotor Hamiltonian matrix in a corresponding basis set of Wigner D-functions $D_{m_1,k_1}^{j_1 *}(\alpha_1, \beta_1, \gamma_1)$. These states are labeled by quantum numbers $\{j_1 m_1 \tau_1\}$ where $j_1$ and $m_1$ represent angular momentum of the first molecule and its projection onto the axis of quantization (defined below). The index $\tau_1 = k_{\mathrm{A}} - k_C$ replaces $k_{\mathrm{A}}$ and $k_C$ that represent projections of $j_1$ onto the principal axis of inertia with smallest and largest values of rotational constants, respectively.

The rotations of a linear rotor are described by polar angles $(\theta, \varphi)$ and its rotational eigenstates are represented by spherical harmonics $\mathrm{Y}_{m_2}^{j_2}(\theta, \varphi)$. Or, for convenience, one can use two of the three Euler angles:

$$\psi_{j_2 m_2}(\Lambda_2) = \mathrm{Y}_{m_2}^{j_2}(\beta_2, \alpha_2) \tag{2}$$

where $\Lambda_2 = (\alpha_2, \beta_2, \gamma_2)$ with the last Euler angle fixed at zero ($\gamma_2 = 0$), while $j_2$ and $m_2$ represent angular momentum of the second molecule and its projection onto the axis of quantization. Then, the coupled states of asymmetric-top rotor + linear rotor can be expressed using Clebsch-Gordan (CG) coefficients $C^{j,m}_{j_1 m_1, j_2 m_2}$ as follows:

$$\Psi_{nm}(\Lambda_1,\Lambda_2) = \sqrt{\frac{2j_1+1}{8\pi^2}} \sum_{m_1=-j_1}^{+j_1} C^{j,m}_{j_1,m_1,j_2,m-m_1} \left( \sum_{k=-j_1}^{+j_1} b_k^{\tau_1} D^{j_1 *}_{m_1,k}(\Lambda_1) \right) \times Y^{j_2}_{m-m_1}(\Lambda_2) \qquad (3)$$

Here $m$ is projection of total angular momentum $j$ of the molecule-molecule system onto the axis of quantization while $n$ is used as a composite index to label the total set of quantum numbers for the system, $n = \{j, m, j_1, \tau_1, j_2\}$. The CG coefficients are nonzero only if $m = m_1 + m_2$ and $|j_1 - j_2| \leq j \leq j_1 + j_2$.

Time evolution of the rotational wavefunction of the system is described by expansion over a set of eigenstates:

$$\psi(\Lambda_1,\Lambda_2,t) = \sum_{nm} a_{nm}(t)\Psi_{nm}(\Lambda_1,\Lambda_2)\exp\{-\mathrm{i}E_n t\} \qquad (4)$$

where $a_{nm}(t)$ is a set of corresponding probability amplitudes that are time-dependent, and the exponential phase factors are included to simplify solution in the asymptotic range. Substitution of this expansion into the time-dependent Schrodinger equation and the transformation of wavefunctions into the rotating frame tied to the molecule-molecule vector $\vec{R}$ (used as a quantization axis in this body-fixed reference frame[49] leads to the following set of coupled equations for time-evolution of probability amplitudes:

$$\dot{a}_{mn''} = -i \sum_{n'} a_{n'm} M^{n''}_{n'}(R)\, e^{i\varepsilon^{n''}_{n'} t}$$
$$-\dot{\Phi}\left[a_{n'',m-1}\sqrt{j''(j''+1)-m(m-1)} + a_{n'',m+1}\sqrt{j''(j''+1)-m(m+1)}\right]/2 \qquad (5)$$

Here $\varepsilon^{n''}_{n'} = E_{n''} - E_{n'}$ is energy difference between the final and initial states of the system. The summation in the first term of this equation includes state-to-state transitions $n' \rightarrow n''$ (within each $m$) driven by real-valued, time-independent potential coupling matrix $M^{n''}_{n'}$:

$$M^{n''}_{n'}(R) = \langle \Psi_{n''m}(\Lambda_1,\Lambda_2) | V(R,\Lambda_1\Lambda_2) | \Psi_{n'm}(\Lambda_1,\Lambda_2) \rangle \qquad (6)$$

The potential energy hypersurface $V(R,\Lambda_1,\Lambda_2)$ depends on the intermolecular distance $R$ and orientation of each molecule, $\Lambda_1$ and $\Lambda_2$. Here $\Psi_{n''m}(\Lambda_1,\Lambda_2)$ and $\Psi_{n'm}(\Lambda_1,\Lambda_2)$ represents wavefunctions of final and initial states of the molecule.

The second term in Eq. (5) describes $m \pm 1 \rightarrow m$ transitions (within each $n$) due to the Coriolis coupling effect, driven by rotation of the molecule-molecule vector $\vec{R} = (R, \Phi, \Theta)$ relative to the laboratory-fixed reference frame during the course of collision. A set of spherical polar coordinates $(R, \Phi, \Theta)$ represents classical degrees of freedom in the system. They describe scattering of two collision partners relative to the laboratory-fixed reference frame and the equations for their time-evolution are obtained using the Ehrenfest theorem:[49]

$$\dot{R} = \frac{P_R}{\mu} \tag{7}$$

$$\dot{\Phi} = \frac{P_\Phi}{\mu R^2} \tag{8}$$

$$\dot{P}_R = -\sum_{n'}\sum_{n''} e^{i\varepsilon_{n'}^{n''}t} \sum_m \frac{\partial M_{n'}^{n''}}{\partial R} a_{n''m}^* a_{n'm} + \frac{P_\Phi^2}{\mu R^3} \tag{9}$$

$$\begin{aligned}\dot{P}_\Phi = -i\sum_{n'}\sum_{n''} e^{i\varepsilon_{n'}^{n''}t} \sum_m M_{n'}^{n''} \\ \times \Big[a_{n''m-1}^* a_{n'm}\sqrt{j''(j''+1)-m(m-1)} \\ + a_{n''m+1}^* a_{n'm}\sqrt{j''(j''+1)-m(m+1)} \\ - a_{n''m}^* a_{n'm-1}\sqrt{j'(j'+1)-m(m-1)} \\ - a_{n''m}^* a_{n'm+1}\sqrt{j'(j'+1)-m(m+1)}\Big]/2\end{aligned} \tag{10}$$

It appears that only the equations for $R$, $\Phi$ and their conjugate momenta $P_R$, $P_\Phi$ are needed. Since the trajectory is planar, one can restrict consideration to the equatorial plane $\Theta = \pi/2$ with $\dot{\Theta} = 0$.[49] Note that classical orbital angular momentum $\dot{\Phi}(t)$ drives Coriolis transitions in the quantum equations of motion, Eq. (5), while the quantum probability amplitudes $a_{nm}(t)$ create a mean-field potential in the classical equations of motion, Eqs. (9-10), providing a link between quantum and classical degrees of freedom. It was demonstrated that the total energy, which is the sum of rotational (quantum) and translational (classical), is conserved along these mixed quantum/classical trajectories.[49,50] The trajectories are propagated for various initial values of the orbital angular momentum of two collision partners $\ell$, and the values of probability amplitudes $a_{nm}(t)$ at the final moment of time are used to determine state-to-state transition probabilities. Those are summed over final and averaged over initial degenerate states to obtain cross sections $\sigma_{n' \rightarrow n''}$ for transitions between non-degenerate states of the system.

In principle, matrix elements of Eq. (6) can be computed by a four-dimensional numerical quadrature:

$$M_{n'}^{n''}(R) = 2\pi \int_0^{\pi} \sin\beta_1 \, d\beta_1 \int_0^{2\pi} d\gamma_1 \times \int_0^{2\pi} d\alpha_2 \int_0^{\pi} \sin\beta_2 \, d\beta_2 \, V(R,\beta_1,\gamma_1,\alpha_2,\beta_2)\Psi_{n''}^{*}(\Lambda_1\Lambda_2)\Psi_{n'}(\Lambda_1\Lambda_2) \quad (11)$$

The factor of $2\pi$ come from the analytical integration over $\alpha_1$. This can be done because potential energy of the system depends only on the relative orientations of two molecules, given by the difference $\alpha_2 - \alpha_1$. One can set $\alpha_1 = 0$, which makes $V(R,\beta_1,\gamma_1,\alpha_2,\beta_2)$ independent of $\alpha_1$. In practice, the multi-dimensional quadrature is numerically expensive. It is better to expand $V(R,\beta_1,\gamma_1,\alpha_2,\beta_2)$ over a set of suitable angular functions $\tau_{\lambda_1\mu_1\lambda_2\lambda}(\beta_1,\gamma_1,\alpha_2,\beta_2)$ with $R$-dependent expansion coefficients $\upsilon_{\lambda_1\mu_1\lambda_2\lambda}(R)$ obtained by projecting $V$ onto these expansion functions at each value of $R$ within a predefined grid. These projections are also computed by numerical quadrature, but the number of expansion functions is much smaller than the number of individual matrix elements. At each value of $R$, the potential can be represented analytically:

$$V(R,\beta_1,\gamma_1,\alpha_2,\beta_2) = \sum_{\lambda_1\mu_1\lambda_2\lambda} \upsilon_{\lambda_1\mu_1\lambda_2\lambda}(R)\tau_{\lambda_1\mu_1\lambda_2\lambda}(\beta_1,\gamma_1,\alpha_2,\beta_2) \quad (12)$$

For an asymmetric-top rotor + linear rotor system a suitable set of functions is given by:

$$\tau_{\lambda_1\mu_1\lambda_2\lambda}(\beta_1,\gamma_1,\alpha_2,\beta_2) = \sqrt{\frac{2\lambda_1+1}{4\pi}} \sum_{\eta=-\min(\lambda_1,\lambda_2)}^{+\min(\lambda_1,\lambda_2)} \mathrm{C}_{\lambda_1,\eta,\lambda_2,-\eta}^{\lambda,0} \times \left[\mathrm{D}_{\eta,\mu_1}^{\lambda_1 *}(\Lambda_1) + (-1)^{\lambda_1+\mu_1+\lambda_2+\lambda}\mathrm{D}_{\eta,-\mu_1}^{\lambda_1 *}(\Lambda_1)\right]\mathrm{Y}_{-\eta}^{\lambda_2}(\Lambda_2) \quad (13)$$

which uses spherical harmonics $\mathrm{Y}_{\eta}^{\lambda}(\beta_2,\alpha_2)$ with the last Euler angle fixed at zero ($\gamma_2 = 0$), Wigner D-functions $D_{m_1,k}^{j_1 *}(\alpha_1,\beta_1,\gamma_1)$ and CG coefficients (see above). The meaning of indexes $\lambda_1$ ($\mu_1$), $\lambda_2$ and $\lambda$ are analogues to angular momenta for the molecule one (its projection onto symmetry axis), the molecule two, and the entire system, respectively. Substitution of Eq. (13) into Eq. (12), and then into Eq. (6) will give us the following expression:

$$M_{n'}^{n''}(R)$$

$$= \sqrt{\frac{2j_1''+1}{8\pi^2}}\sqrt{\frac{2j_1'+1}{8\pi^2}}\sum_{m_1'=-j_1'}^{+j_1'} \mathrm{C}_{j_1',m_1',j_2',m-m_1'}^{j',m}\sum_{m_1''=-j_1''}^{+j_1''} \mathrm{C}_{j_1'',m_1'',j_2'',m-m_1''}^{j'',m}\sum_{\lambda_1\mu_1\lambda_2\lambda} \upsilon_{\lambda_1\mu_1\lambda_2\lambda}(R)\sqrt{\frac{2\lambda_1+1}{4\pi}}$$

$$\times \sum_{\eta=-\min(\lambda_1,\lambda_2)}^{+\min(\lambda_1,\lambda_2)} \mathrm{C}_{\lambda_1,\eta,\lambda_2,-\eta}^{\lambda,0}\sum_{k_1''=-j_1''}^{+j_1''}\sum_{k_1'=-j_1'}^{+j_1'} b_{k_1''}^{\tau_1''} b_{k_1'}^{\tau_1'}\left\langle \mathrm{Y}_{m-m_1''}^{j_2''}(\Lambda_2)\middle|\mathrm{Y}_{-\eta}^{\lambda_2}(\Lambda_2)\middle|\mathrm{Y}_{m-m_1'}^{j_2'}(\Lambda_2)\right\rangle$$

$$\times\left[\left\langle \mathrm{D}_{m_1'',k_1''}^{j_1''*}(\Lambda_1)\middle|\mathrm{D}_{\eta,\mu_1}^{\lambda_1*}(\Lambda_1)\middle|\mathrm{D}_{m_1',k_1'}^{j_1'*}(\Lambda_1)\right\rangle + (-1)^{\lambda_1+\mu_1+\lambda_2+\lambda}\left\langle \mathrm{D}_{m_1'',k_1''}^{j_1''*}(\Lambda_1)\middle|\mathrm{D}_{\eta,-\mu_1}^{\lambda_1*}(\Lambda_1)\middle|\mathrm{D}_{m_1',k_1'}^{j_1'*}(\Lambda_1)\right\rangle\right] \quad (14)$$

For the integrals of spherical harmonics and Wigner D-functions, the following properties can be employed:[51]

$$\left\langle \mathrm{Y}_{m_3}^{l_3}\middle|\mathrm{Y}_{m_2}^{l_2}\middle|\mathrm{Y}_{m_1}^{l_1}\right\rangle = \sqrt{\frac{(2l_1+1)(2l_2+1)}{4\pi(2l_3+1)}}\mathrm{C}_{l_1,0,l_2 0}^{l_3,0}\mathrm{C}_{l_1,m_1,l_2,m_2}^{l_3,m_3} \quad (15)$$

$$\left\langle \mathrm{D}_{m_3 m_3'}^{l_3*}\middle|\mathrm{D}_{m_2 m_2'}^{l_2*}\middle|\mathrm{D}_{m_1 m_1'}^{l_1*}\right\rangle = \frac{8\pi^2}{2l_3+1}\mathrm{C}_{l_1,m_1,l_2,m_2}^{l_3,m_3}\mathrm{C}_{l_1,m_1',l_2,m_2'}^{l_3,m_3'} \quad (16)$$

By substituting the integrals in Eq. (14) with the expressions Eq. (15) and Eq. (16) we obtain the following expression:[51]

$$M_{n'}^{n''}(R) = \sqrt{\frac{2j_1''+1}{8\pi^2}}\sqrt{\frac{2j_1'+1}{8\pi^2}}\sum_{m_1'=-j_1'}^{+j_1'} \mathrm{C}_{j_1',m_1',j_2',m-m_1'}^{j',m}\sum_{m_1''=-j_1''}^{+j_1''} \mathrm{C}_{j_1'',m_1'',j_2'',m-m_1''}^{j'',m}$$

$$\times \sum_{\lambda_1\mu_1\lambda_2\lambda} \upsilon_{\lambda_1\mu_1\lambda_2\lambda}(R)\sqrt{\frac{2\lambda_1+1}{4\pi}}\sum_{k_1''=-j_1''}^{+j_1''}\sum_{k_1'=-j_1'}^{+j_1'} b_{k_1''}^{\tau_1''} b_{k_1'}^{\tau_1'}\sum_{\eta=-\min(\lambda_1,\lambda_2)}^{+\min(\lambda_1,\lambda_2)} \mathrm{C}_{\lambda_1,\eta,\lambda_2,-\eta}^{\lambda,0}\sqrt{\frac{(2j_2'+1)(2\lambda_2+1)}{4\pi(2j_2''+1)}}$$

$$\times \mathrm{C}_{j_2',0,\lambda_2 0}^{j_2'',0}\mathrm{C}_{j_2',m-m_1',\lambda_2-\eta}^{j_2'',m-m_1''}\frac{8\pi^2}{2j_1''+1}\mathrm{C}_{j_1',m_1',\lambda_1,\eta}^{j_1'',m_1''}\left[C_{j_1',k_1',\lambda_1,\mu_1}^{j_1'',k_1''} + (-1)^{\lambda_1+\mu_1+\lambda_2+\lambda} C_{j_1',k_1',\lambda_1,-\mu_1}^{j_1'',k_1''}\right] \quad (17)$$

In general, Clebsch-Gordan coefficients are non-zero only if $m = m_1 + m_2$ and $|j_1 - j_2| \le j \le j_1 + j_2$. Incorporating these properties of CG coefficients[51] we obtain the final state-to-state transition matrix element as follows:

$$M_{n'}^{n''}(R) = \sqrt{\frac{2j_1'+1}{2j_1''+1}}\sqrt{\frac{2j_2'+1}{2j_2''+1}}\sum_{\lambda_1\mu_1\lambda_2\lambda} \upsilon_{\lambda_1\mu_1\lambda_2\lambda}(R)\sqrt{\frac{2\lambda_1+1}{4\pi}}\sqrt{\frac{2\lambda_2+1}{4\pi}} C_{j_2',0,\lambda_2,0}^{j_2'',0}$$

$$\times \sum_{m_1'=-j_1'}^{+j_1'} \mathrm{C}_{j_1',m_1',j_2',m-m_1'}^{j',m}\sum_{\eta=-\min(\lambda_1,\lambda_2)}^{+\min(\lambda_1,\lambda_2)} \mathrm{C}_{j_1'',m_1'-\eta,j_2'',m-(m_1'-\eta)}^{j'',m} C_{\lambda_1,\eta,\lambda_2,-\eta}^{\lambda,0}\mathrm{C}_{j_1',m_1'-\eta,\lambda_1,\eta}^{j_1'',m_1'}$$

$$\times C^{j_2'',m-m_1'}_{j_2',m-(m_1'-\eta),\lambda_2,-\eta} \sum_{k_1''=-j_1''}^{+j_1''} \sum_{k_1'=-j_1'}^{+j_1'} b^{\tau_1''}_{k_1''} b^{\tau_1'}_{k_1'} \left[ C^{j_1'',k_1''}_{j_1',k_1',\lambda_1,\mu_1} + (-1)^{\lambda_1+\mu_1+\lambda_2+\lambda} C^{j_1'',k_1''}_{j_1',k_1',\lambda_1,-\mu_1} \right] \tag{18}$$

In order to check the correctness of Eq. (18), and to assess the accuracy of PES expansion, we computed matrix elements at two values of the molecule-molecule distance in the interaction region, $R$ = 7.0 and 8.0 Bohr, for 14 states in the basis that include combinations of ground and excited states of both collisions partners: $(j_{1_{k_A k_C}} j_2) = (0_{00}0)$, $(1_{11}0)$, $(2_{02}0)$, $(2_{11}0)$, $(2_{20}0)$, $(3_{13}0)$, $(3_{22}0)$, $(0_{00}2)$, $(1_{11}2)$, $(2_{02}2)$, $(2_{11}2)$, $(2_{20}2)$, $(3_{13}2)$, $(3_{22}2)$. Note that this list includes all symmetries of $H_2O$ + $H_2$ system, namely, $p$-$H_2O$ + $p$-$H_2$, $p$-$H_2O$ + $o$-$H_2$, $o$-$H_2O$ + $p$-$H_2$, and $o$-$H_2O$ + $o$-$H_2$. Analytical representation of the PES included 83 expansion terms with $\lambda_1$ and $\mu_1$ up to 9 and 6, respectively, $\lambda_2$ up to 6 and the overall $\lambda$ up to 11, respectively. We employed numerical quadrature to obtain the expansion coefficients $v_{\lambda_1\mu_1\lambda_2\lambda}(R)$ as required in Eq. (12), and to calculate the same matrix elements through direct integration using Eq. (11). Equidistant grids with 40 points were used for $\alpha$ and $\gamma$, and a Gauss-Legendre method with 20 points was used for $\beta$. The values of computed matrix elements in this test ranged from $10^{-4}$ to $10^{2}$ $cm^{-1}$, while the difference between the two methods of calculations (direct integration vs PES expansion) was found to be on the order of $10^{-10}$ $cm^{-1}$, from which we conclude that both methods work as expected. Also, this test permitted us to compare numerical costs of the two methods and showed that the PES expansion method gives a significant computational advantage. For the specified subset of rotational states, its CPU cost is eleven times lower (the method is faster) compared to the direct integration method.

## III. DETAILS OF CALCULATIONS

In the calculations presented below the rotational basis set of water includes states up to $j_1 = 28$ for both $p$-$H_2O$ and $o$-$H_2O$ with energies reaching 7907 and 7996 $cm^{-1}$, respectively. These are combined with the states of $p$-$H_2$ and $o$-$H_2$ up to $j_2 = 10$, that have energies up to 6694 $cm^{-1}$. Combinations of these states give 1032 energetically non-degenerate channels for $o$/$p$-target + $p$-projectile and 897 channels for $o$/$p$-target + $o$-projectile, with total rotational energy below 8000 $cm^{-1}$. All possible combinations of angular momenta of two collision partners $j_1$ and $j_2$ were included up to the total $j = 38$ for $o$/$p$-target + $p$-projectile and $j = 37$ for $o$/$p$-target + $o$-projectile, with all possible values of quantum number $m$ for its projection onto the axis of quantization $(-j \leq$ $m \leq +j$) which resulted in 206350 states in the $p$-$p$ case, 205649 states in the $p$-$o$ case, 206812 states in the $o$-$p$ case, and 205211 states in the $o$-$o$ case. This choice of the rotational basis set led to the state-to-state transition matrix amounting to approximately half a billion non-zero matrix elements in total. This entire matrix was computed and analyzed at two values of the molecule-molecule distance in the interaction region, $R$ = 6.03 and 7.11 Bohr. Subsequently, based on this analysis, the matrix was truncated, retaining only the elements with magnitudes above 1 cm$^{-1}$, which resulted in 14580217, 14610968, 14628390 and 14575564 matrix elements (transitions) retained for $p$-$p$, $p$-$o$, $o$-$p$ and $o$-$o$ symmetries of $H_2O$ + $H_2$ system, respectively. Next, the values of these retained matrix elements (the truncated matrix) were computed on a grid of 63 points along $R$ that cover the range $3.0 \leq R \leq 30$ Bohr using logarithmic spacings to make the grid denser at short range and progressively sparser as the molecule-molecule separation approaches the asymptotic region. Finally, these data were interpolated using cubic spline. The calculations of state-to-state transition matrix elements were carried out in parallel using 5 nodes of HPC *Raj* at Marquette University (AMD Rome 2 GHz processors, 512 GB memory) where each node has 128 processors, leading to overall 640 processors per job. Four different matrices were computed for the combinations involving *para*- and *ortho*-states of target and projectile, as described above. Each of these matrix computations took about 96 hours (4 wall-clock days), with the total cost close to 250,000 CPU hours.

The calculations of collision dynamics were conducted using AT-MQCT version of theory[52], in which the propagation of Eqs. (5) for quantum degrees of freedom is decoupled from the propagation of Eqs. (7-10) for classical degrees of freedom. This approach is known to give accurate results for $H_2O$ + $H_2$.[52] An adaptive-step-size predictor method ADAPTOL was employed for time-propagation[52], with the tolerance parameter $\epsilon$ = 10$^{-3}$ determined by convergence studies. This method uses larger time steps in the asymptotic region of the PES, and then reduces the time step in the interaction region, down to 10 a.u. Initial conditions for MQCT trajectories were sampled randomly using a Monte-Carlo approach with maximum value of impact parameter $b_{max} = 25$ Bohr for the lowest collision energy and $b_{max} = 15$ Bohr for all higher energies, that correspond to orbital angular momentum quantum number $\ell_{max} = 27$ and 285, respectively. Overall, we conducted MQCT calculations for ten collision energies: $U$ = 20.00, 41.28, 84.00, 170.47, 346.41, 703.89, 1430.0, 2906.3, 5906.0 and 12000 cm$^{-1}$. For each energy, 200 randomly sampled MQCT trajectories were propagated using 64 processors per trajectory. These randomly sampled trajectories were used to compute state-to-state transition cross sections as described in our earlier papers.[31,40] We found that 200 trajectories were sufficient to obtain cross sections converged to within ~ 1% percent of their values for most transitions. However, for transitions with larger values of $j_1$ and $j_2$ the convergence was estimated to be within 10%. Calculations were carried out for the initial states that combine 100 states of water (of each symmetry) with eleven considered states of $H_2$ (both symmetries). On average, calculations for 100 initial states of water combined with one initial state of $H_2$, took about 480 minutes at one collision energy, which translates to around 1600 hours of wall clock time, or 102400 CPU hours dedicated to the calculations. The overall numerical cost of MQCT trajectories for ten collision energies, four symmetries of the system, and all the initial states of $H_2O$ + $H_2$ was about 22 million CPU hours on *Raj* cluster at Marquette (see above). This computational cost is a factor of four higher than that of our recent calculations for $H_2O$ + $H_2O$ system,[45] which is partially due to more collision energies (ten here vs six in that work) and partially due to technical issues that decreased the performance of our cluster.

For each individual transition in *o*/*p*-$H_2O$ + *o*/*p*-$H_2$ we computed state-to-state transition cross sections for both quenching and excitation directions of each process. These cross sections satisfy the principle of microscopic reversibility if the following conditions is fulfilled:

$$(2j_1' + 1)(2j_2' + 1)\sigma_{n'\to n''} = (2j_1'' + 1)(2j_2'' + 1)\sigma_{n''\to n'} \quad (19)$$

Here and below $n = (j_{1_{k_A k_C}} j_2)$ labels non-degenerate states of the molecule-molecule system, while prime and double prime indexes define the initial and final states, say $n'' \to n'$ is for quenching and $n' \to n''$ for excitation. It should be noted that the mean-field Ehrenfest method does not automatically satisfy the principle of microscopic reversibility. To overcome this drawback a so-called Billing correction is applied to computed cross sections,[44,53–55] which permits to satisfy Eq (19) approximately. One can also check for possible deviations by doing calculations of both excitation and quenching cross sections. In Fig. 1 we present a comparison of the left- *vs* right-hand side of Eq. (19) with Billing correction applied, at three collision energies for 14388 transitions in $H_2O$ + $H_2$ system, which includes 100 initial states of *p*-$H_2O$ combined with three initial states for *p*-$H_2$ ($j_2' = 0, 2, 4$). The other three symmetries of $H_2O$ + $H_2$ system were analyzed in a similar way and all the results are presented in Supplementary Information, Fig. S1. Large differences between the left- and right-hand sides of Eq. (19) would indicate possible flaws in the physical assumptions of the method and/or in numerical aspect of calculations. However, the data presented in Fig. 1 and Fig. S1 demonstrate that, in general, the microscopic reversibility is approximately satisfied with deviations remaining relatively small (within a factor of 2) for most transitions across several orders of magnitude range of cross section values. Larger deviations are found for smaller cross sections and at lower collision energies, which is expected because an approximate trajectory-based method may indeed become somewhat less accurate at lower energies. Moreover, we looked separately at the data that corresponds to the initial $H_2$ ($j_2' = 0$), $H_2$ ($j_2' = 2$) and $H_2$ ($j_2' = 4$) and found that the deviations from the diagonal line are similar and does not get worse as the initial rotational excitation of $H_2$ is raised.

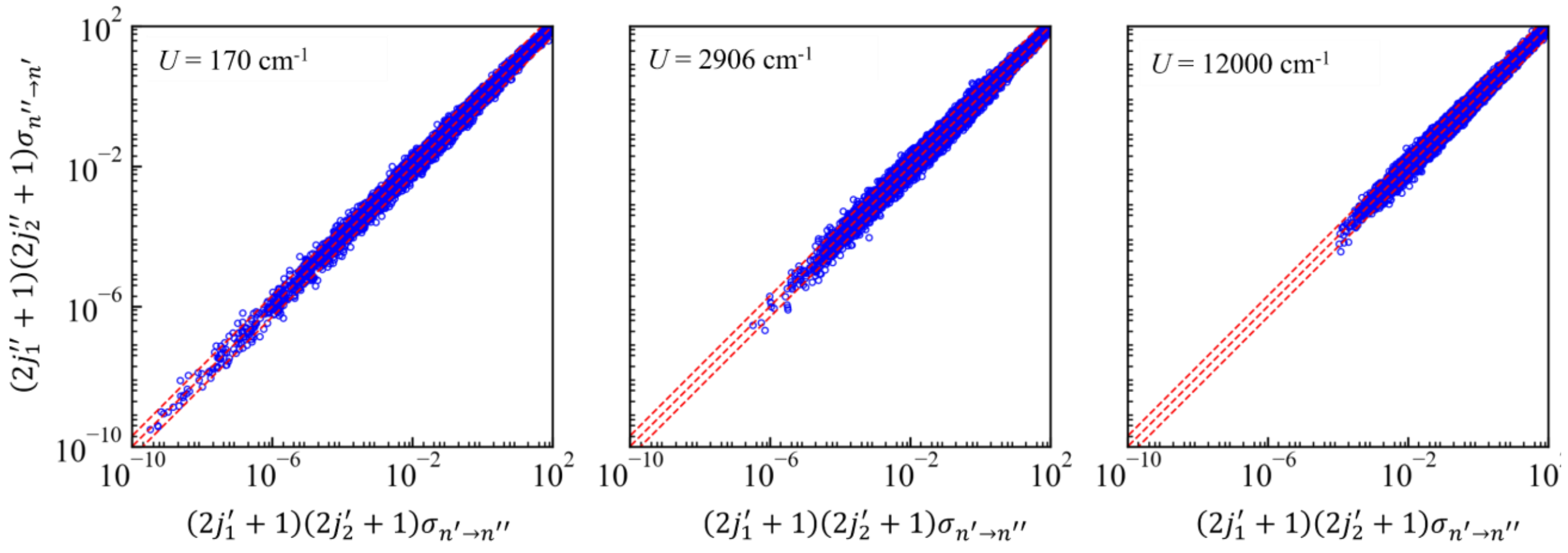

**Figure 1:** The comparison of MQCT state-to-state transition cross sections (in Å$^2$) for quenching and excitation directions of 14388 individual transitions in *p*-$H_2O$ + *p*-$H_2$ system at three values of collision energy *U*. The deviation of datapoints from the diagonal line indicates the departure from the principle of microscopic reversibility. The factor of 2 difference is shown by red dashed lines.

In order to understand the origin of these differences, we selected fifteen transitions characterized by some of the largest differences between quenching and excitation cross sections in Fig. 2, at the lowest collision energy $U = 170$ cm$^{-1}$. The selected transitions were: $9_{91}0 \to 7_{71}2$, $12_{93}2 \to 12_{84}2$, $12_{93}2 \to 11_{84}2$, $13_{410}2 \to 12_{210}0$, $8_{44}2 \to 6_{60}2$, $6_{06}2 \to 2_{02}2$, $13_{410}2 \to 12_{210}2$, $12_{75}0 \to 13_{59}0$, $13_{410}2 \to 11_{84}0$, $13_{410}2 \to 10_{82}2$, $10_{37}2 \to 6_{51}2$, $8_{08}2 \to 4_{40}2$,

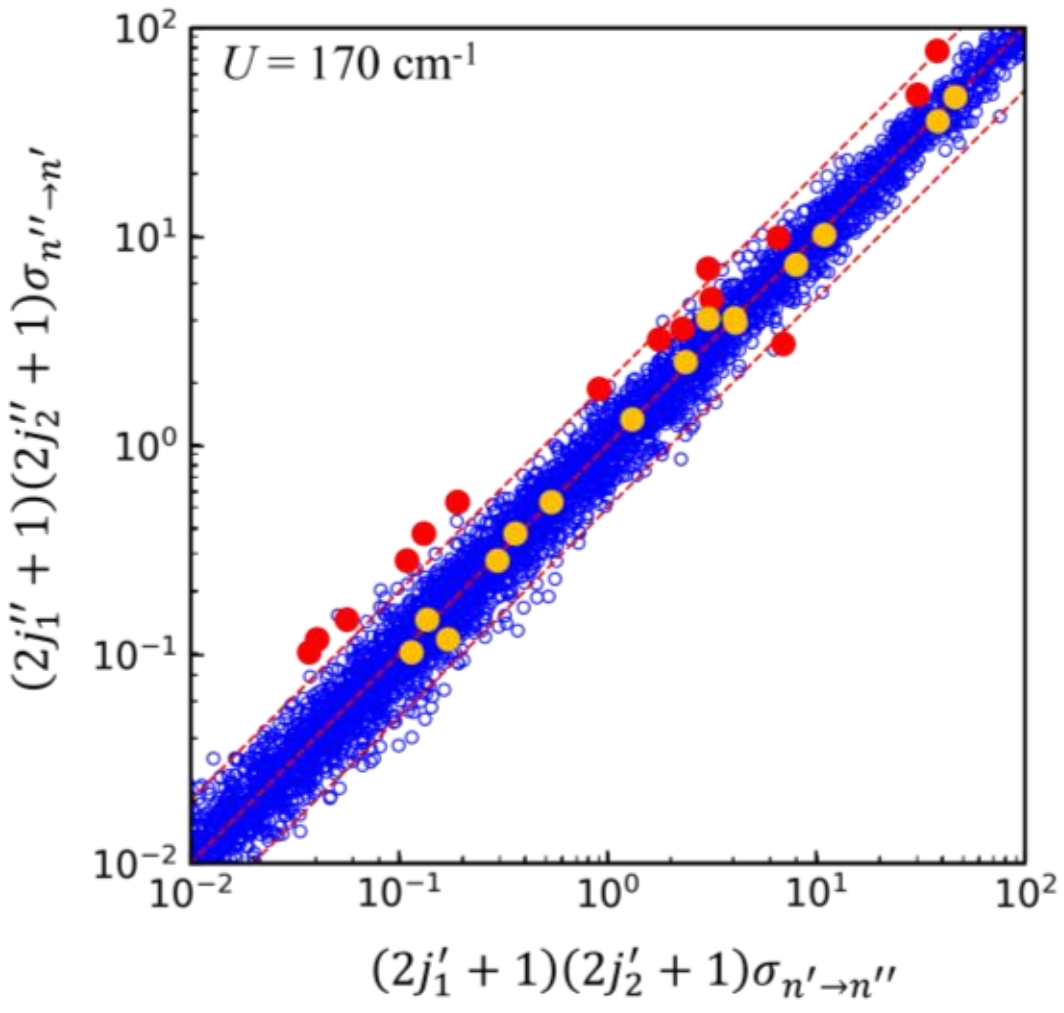

**Figure 2:** The effect of initial sampling on excitation and quenching cross sections for fifteen individual transitions in in *p*-$H_2O$ + *p*-$H_2$ system at $U$ = 170 cm$^{-1}$, as explained in the text. The datapoints obtained by random sampling of initial conditions (red symbols) show larger deviations from the principle of microscopic reversibility. Regular sampling for the same fifteen transitions (yellow symbols) reduces the deviation significantly.

$15_{214}2 \rightarrow 14_{113}2$, $13_{410}2 \rightarrow 12_{111}2$ and $16_{016}2 \rightarrow 17_{117}0$. These 15 datapoints, highlighted in red in Fig. 2, were then recalculated using a regular sampling of initial conditions (without random sampling, with trajectories initiated for all initial values of quantum numbers $j$, $m$ and $\ell$ varied through the following ranges: $|j_1 - j_2| \leq j \leq j_1 + j_2$, $-j \leq m \leq +j$ and $0 \leq \ell \leq \ell_{max}$). The results were re-plotted as yellow symbols in Fig. 2.

From Fig. 2 we see that yellow symbols are found much closer to the diagonal line, compared to red symbols, which means that larger deviations of the datapoints from the principle of microscopic reversibility originate mostly in the random sampling of initial conditions for the calculations of excitation and quenching cross sections. This also means that, in principle, one can improve the results of MQCT calculations if needed, for an extra cost, by either running more randomly sampled trajectories for transitions with larger values of $j_1'$, or by refraining from the random sampling at all (and propagating all possible trajectories as we did in this example), which would be practical only for small values of $j_1'$ or for a small number of transitions.

Here we decided not to do more calculations, simply because the agreement between the excitation and quenching is already within the desired range and remains consistent for all symmetries of $H_2O + H_2$ system through a broad range of collision energies and for all transitions we studied (close to 118 thousand transitions in total, see Fig. S1). In order to obtain rate coefficients (that must satisfy the principle microscopic reversibility) we computed the weighted average of cross sections for excitation and quenching:

$$\tilde{\sigma}_{n'n''} = \frac{(2j_1' + 1)(2j_2' + 1)\sigma_{n' \to n''} + (2j_1'' + 1)(2j_2'' + 1)\sigma_{n'' \to n'}}{2} \tag{20}$$

and used these values to calculate state-to-state transition rate coefficients for both quenching and excitation processes, $k_{n'' \to n'}(T)$ and $k_{n' \to n''}(T)$, respectively, by integrating $\tilde{\sigma}_{n'n''}$ over the Boltzmann-Maxwell distribution of collision energies as described elsewhere.[44] For this integration we constructed a cubic spline through ten energy points where these cross sections were computed (see above) and extrapolated these data towards the excitation threshold at low collision energy where cross section goes to zero, and towards high collision energies, as described elsewhere.[44]

## IV. COMPARISON WITH FULL QUANTUM RESULTS

The rate coefficients for collisions between $H_2O$ and $H_2$ molecules were determined by full-quantum calculations in several earlier studies.[56–59] The most comprehensive work reported by Daniel et al.[59] employed an accurate coupled-channel method of MOLSCAT package[60] and covered all four symmetries of the $H_2O + H_2$ system. Those data are available from BASECOL database.[61] For *p*-$H_2O$ + *p*-$H_2$ symmetry they include rate coefficients for quenching of 44 rotationally excited states of water with rotational levels up to $j_1 = 11$, collided with hydrogen in its ground rotational state, $j_2 = 0$. Additionally, for collisions of *p*-$H_2O$ with *p*-$H_2$ in its excited rotational state $j_2 = 2$ the rate coefficients are available for quenching of 19 excited states of water, up to $j_1 = 7$. For *p*-$H_2O$ + *o*-$H_2$ symmetry, rate coefficients were computed for quenching of the 19 excited states of water up to $j_1 = 7$ collided with the ground state $j_2 = 1$ of hydrogen, and for 9 excited states of water up to $j_1 = 4$ collided with the excited state $j_2 = 3$ of hydrogen. Likewise, for *o*-$H_2O$ + *p*-$H_2$ symmetry, they have provided quenching rate coefficients for the excited 44 levels of water collided with both $j_2 = 0$ and 2 of hydrogen, and additionally for the 9 excited states of ortho-water collided with hydrogen in $j_2 = 4$ state. Finally, regarding the *o*-$H_2O$ + *o*-$H_2$ symmetry, quenching rate coefficients were reported for the excited 44 levels of water collided with hydrogen in $j_2 = 1$ and for the excited 4 states of water collided with $j_2 = 3$ state of hydrogen.

Overall, 8925 individual state-to-state transition rate coefficients are available from the work of Daniel et al.[59] All details are given in the Supplemental Information (please see Table S1). In our pursuit to check the accuracy of new rate coefficients computed here using MQCT, we conducted a rigorous assessment by comparing our results with those of Daniel et al.[59] for all the 8925 transitions reported in their work. Here we used the same PES as employed in the full quantum calculations.[62]

Figure 3 illustrates the comparison of MQCT rate coefficients for 2360 individual state-to-state transitions in the case of *p*-$H_2O$ + *p*-$H_2$, at four temperatures, with the full quantum data of Daniel et al.[59] A similar comparison for three other symmetries of $H_2O + H_2$ system is presented in Figs. S2-S4 of the Supplementary Information. The quenching of excited water states happens simultaneously with quenching, excitation, or elastic scattering of the projectile, namely: 0→0, 0→2, 2→0 and 2→2 transitions in $H_2$. Datapoints that correspond to these processes are shown in Fig. 2 by four different colors, as indicated in the figure legend. The upper row of frames in Fig. 3 gives one-to-one comparison of the rate coefficients (in the units of $cm^3/s$) giving a large picture of qualitative agreement between the two methods, while the lower row of frames reports differences (in %, relative to the full quantum results) giving a more quantitative measure of differences.

From Fig. 3 and Figs. S2-S4 one can see that, overall, the results of MQCT are in good systematic agreement with results of the full quantum coupled-channel calculations, for all transitions considered, through two orders of magnitude range of rate coefficient values, and for a broad range of temperatures. For the most intense transitions with large cross sections the differences between the two methods are within 20%, which is great. As the values of cross sections decrease by an order of magnitude, the differences between the two methods increase but remain within 50%. For the weakest transitions (with cross sections that are two orders of magnitude smaller than those of the strongest transitions), the differences between MQCT and MOLSCAT results remain within 80%. There are only a few data points in Fig. 3 and Figs. S2-4 that indicate differences above 60%, but those correspond to very small cross sections. Green dashed line in the lower row of frames in Fig. 3 and Figs. S2-4 serves as a reference and one can notice that, on average, the values of cross sections computed by MQCT are somewhat smaller than those obtained from full quantum calculations (MOLSCAT).

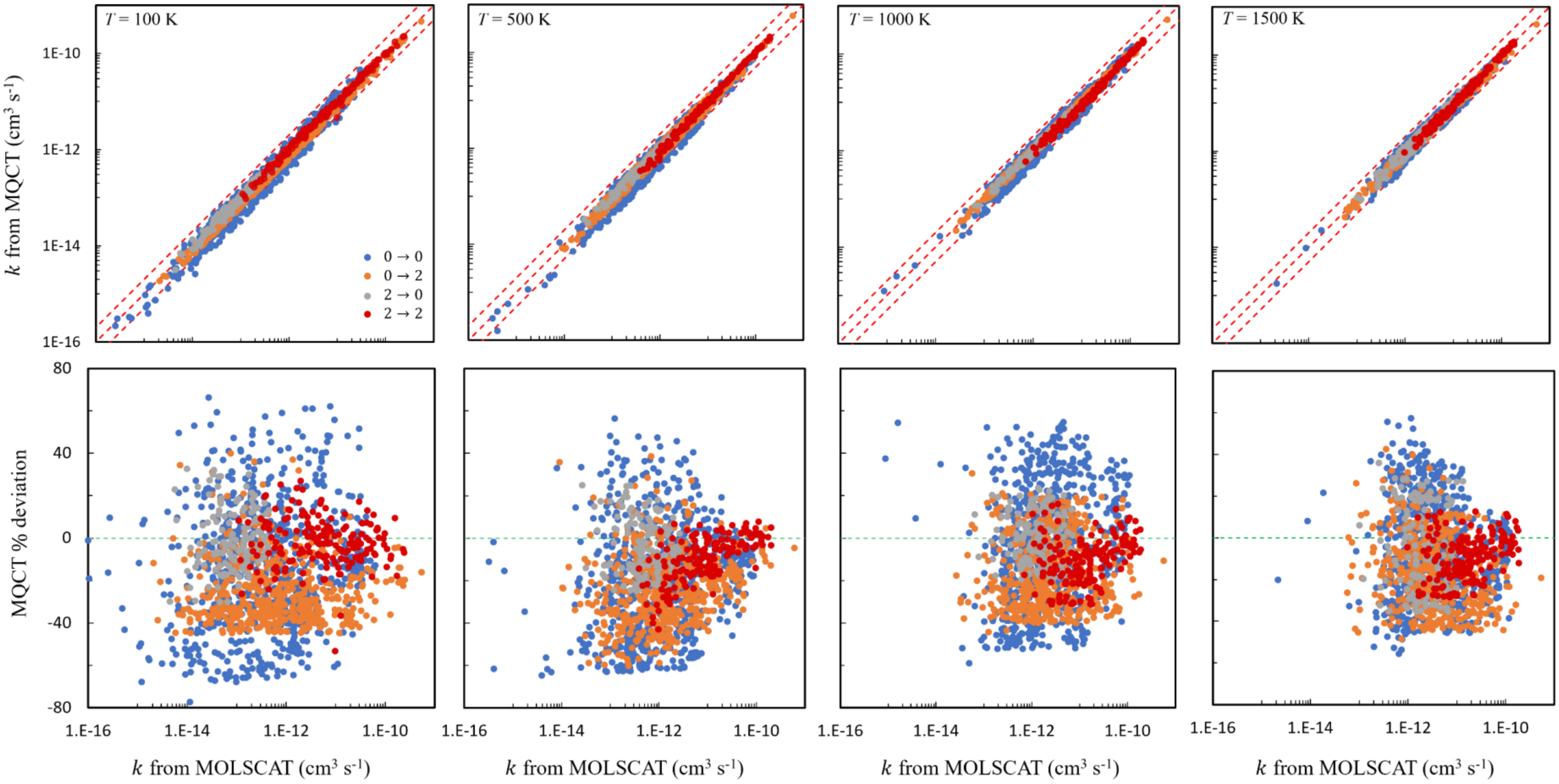

**Figure 3:** Comparison of 2360 state-to-state transition rate coefficients for *p*-$H_2O$ + *p*-$H_2$ collision computed using MQCT (this work) *vs* those predicted by full quantum MOLSCAT calculations.[59] Columns correspond to four values of temperature as indicated in the figure. The upper row of frames gives a one-to-one comparison of rate coefficients, while the lower row of frames presents deviations (in %) of MQCT data relative to MOLSCAT. Color is used to differentiate transitions in the projectile, namely, 0→0, 0→2, 2→0 and 2→2 transitions in $H_2$ are represented by blue, orange, grey and maroon, respectively. Red dashed lines in the upper row represent a factor of 2 difference.

**Table 1:** Average difference and RMS deviation (in %) for the data presented in Fig. 1 for *p*-$H_2O$ + *p*-$H_2$ system, of various transitions in the projectile (0→0, 0→2, 2→0 and 2→2) at four different temperatures.

| Transitions in $H_2$ | T (K) | Average difference (%) | RMS of % deviation |
|---|---|---|---|
| 0 → 0 | 100 | -14 | 33 |
| | 500 | -15 | 31 |
| | 1000 | -10 | 27 |
| | 1500 | -9 | 24 |
| 0 → 2 | 100 | -21 | 24 |
| | 500 | -12 | 21 |
| | 1000 | -6 | 17 |
| | 1500 | -3 | 10 |
| 2 → 0 | 100 | -0.4 | 15 |
| | 500 | -4 | 14 |
| | 1000 | -2 | 12 |
| | 1500 | -6 | 20 |
| 2 → 2 | 100 | -2 | 12 |
| | 500 | -10 | 14 |
| | 1000 | -11 | 15 |
| | 1500 | -9 | 14 |

These differences are further quantified in Table 1 that reports average difference and RMS deviation (in %) for *p*-$H_2O$ + *p*-$H_2$, and in Table S2 of Supplemental Information for three other symmetries of the system. From these tables one can clearly see that the results of MQCT calculations are somewhat smaller than those of the full quantum calculations with MOLSCAT. Interestingly, the differences are notably smaller for transitions where $H_2$ was in the excited rotational state initially. In these cases, the differences tend to increase with temperature. In the cases when *o*/*p*-$H_2$ was in its ground rotational state the differences tend to decrease at high temperature. For example, for the ground state $H_2$, averaged over all transitions, temperatures, and symmetries the average difference between MQCT and MOLSCAT rate coefficients is $-11\%$, while this difference is only −5% in collisions with initially excited $o/p$-$H_2$. The RMS deviations show the same trend, 21% and 11% in these two cases, respectively. Note that for many cells in Table 1 and Table S2 the differences between MQCT and the full quantum MOLSCAT results are within 10%.

These differences can be partially attributed to different sizes of rotational basis sets in two calculations. Namely, in our MQCT calculations the basis set is much larger than in the full quantum study, enabling more pathways for the rotational excitation, which is particularly important for the high energy states of water and high energy collisions (higher temperature). As more high-energy states are populated, the populations of low-energy states tend to decrease (due to norm conservation) leading to smaller cross sections, and we think that this effect explains our observations. To examine the influence of basis set size, we selected from Fig. 2 twenty transitions characterized by some of the largest differences between MQCT and MOLSCAT, at the lowest and highest temperatures, namely: $6_{60}0 \to 5_{51}0$, $7_{62}0 \to 6_{51}0$, $8_{44}0 \to 7_{53}0$, $8_{53}0 \to 7_{62}0$, $8_{53}0 \to 9_{37}0$, $6_{06}0 \to 5_{15}0$, $9_{28}0 \to 8_{35}0$, $7_{53}0 \to 6_{42}0$, $6_{51}0 \to 5_{42}0$, $4_{31}0 \to 4_{13}0$ for $T =$ 100 K, and $5_{33}0 \to 2_{11}0$, $10_{010}0 \to 6_{66}0$, $8_{35}0 \to 8_{26}0$, $6_{60}0 \to 5_{24}0$, $2_{11}0 \to 2_{02}0$, $3_{31}0 \to 2_{22}0$, $4_{31}0 \to 3_{13}0$, $2_{22}0 \to 0_{00}0$, $8_{53}0 \to 6_{24}0$, $9_{28}0 \to 7_{17}0$ for $T =$ 1500 K. These 20 datapoints, highlighted in red in Fig. 4, were then recalculated using a reduced basis set, comparable to that used in the full quantum calculations. Namely, only 45 rotational states of $H_2O$ up to $8_{62}$ combined with two states of hydrogen ($j_2 = 0, 2$) resulting in the total energy up to 1810 cm$^{-1}$ for $H_2O$ + $H_2$ system, were included in the basis in this numerical experiment. These recalculated rate coefficients are shown as yellow symbols in Fig. 4. We see that, at low temperatures, reducing the basis set size improved the agreement between MQCT and MOLSCAT for most of the transitions. At high temperature also, the reduction of basis set size made a similar improvement: The agreement between MQCT and MOLSCAT has improved for all ten examined transitions. In particular, for the transitions where the differences on the order of 60% were observed initially, they decreased to about 40% upon the basis set reduction (see Fig. 4). From this we conclude that the differences between the past MOLSCAT and present MQCT calculations seen in Fig. 2 (and Figs. S2-S4) are partially explained by different sizes of rotational basis sets. Therefore, the larger basis set used in MQCT calculations in this work represents an improvement over the past work.

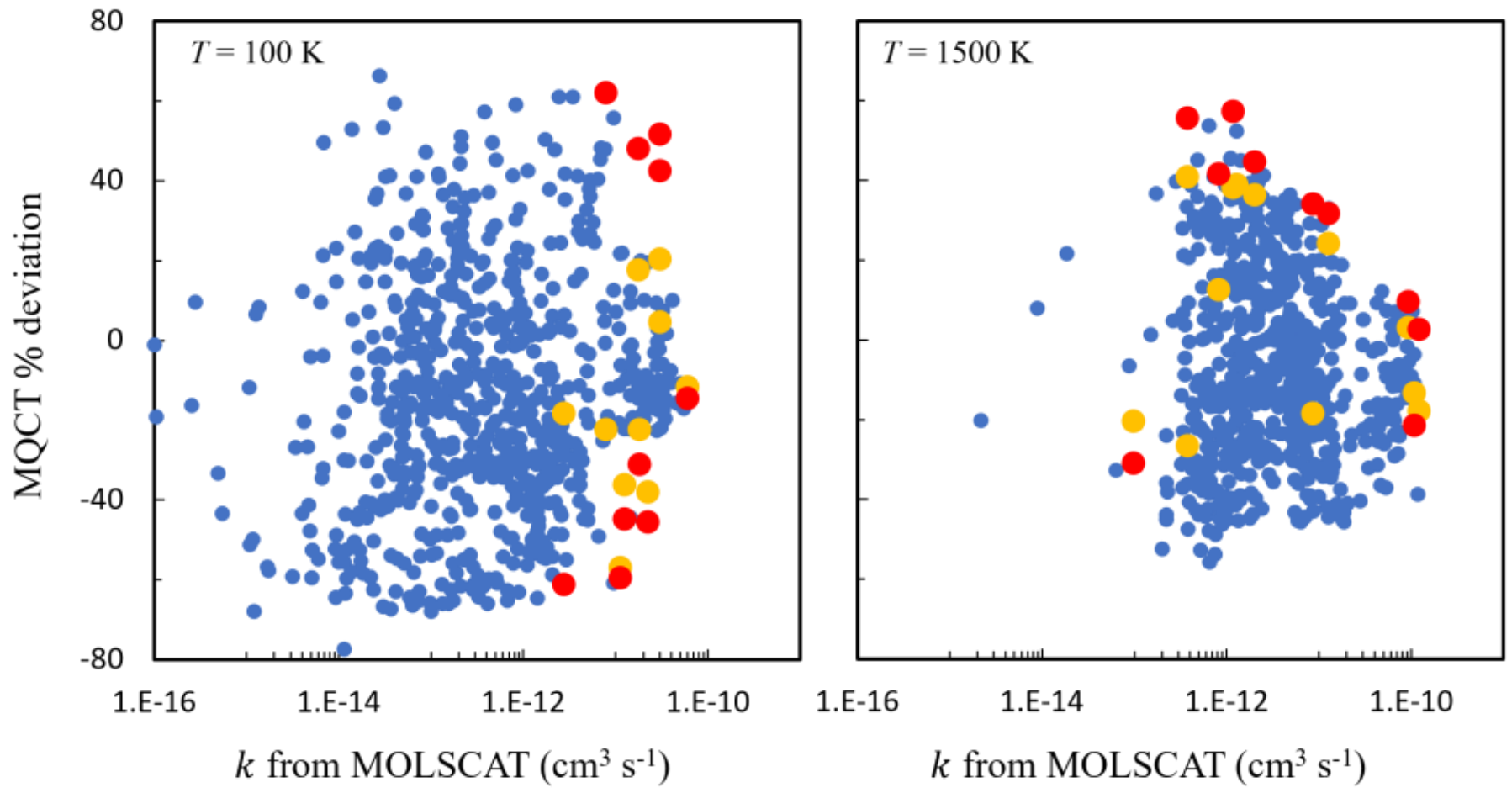

**Figure 4:** The effect of rotational basis set reduction on state-to-state transition rate coefficients in $p$-$H_2O$ + $p$-$H_2$ at low and high temperatures, as indicated in the figure. The datapoints for twenty-two examined transitions with large differences between the past MOLSCAT (smaller basis set) and present MQCT (large basis set) calculations are shown in red. Yellow symbols show results recomputed for the same transitions using MQCT with a reduced basis set.

## V. ROTATIONAL ENERGY TRANSFER IN $H_2O$ + $H_2$ COLLISIONS

Collisional energy transfer is often characterized by a so-called energy transfer kernel,[63–65] or energy transfer function $p(E', E'')$[66,67] that gives probability for the molecule to have internal energy $E''$ after the collision, if before the collision it had the internal energy $E'$. These energy transfer functions are determined from quantum scattering theory[68–71] or classical trajectory simulations[72,73] and are usually described parametrically by an exponential dependence $p = \exp\{-\Delta E/\varepsilon\}$, where $\Delta E = E'' - E'$ while $\varepsilon$ is a fitting coefficient, positive for excitation and negative for quenching, which can also be made energy- or temperature-dependent. In some systems, in order to describe a more complicated profile of collisional energy transfer,[74,75] a double-exponential model may need to be employed, e.g.: $p = a\exp\{-\Delta E/\varepsilon_1\} + (1 - a)\exp\{-\Delta E/\varepsilon_2\}$. Even more detailed models exist,[76] that build a multi-parameter energy transfer function, in which the transition probability depends not only on the initial and final values of energy, but also on the initial $j'$ and final $j''$ values of the angular momentum of the molecule, e.g., $p = \exp\{-\Delta E/\varepsilon\}\exp\{-\Delta j/\gamma\}$, where $\Delta j = j'' - j'$. In what follows, we present a detailed analysis of collisional energy transfer through the dependencies of state-to-state transition cross sections $\sigma_{n'\to n''}$ in $H_2O$ + $H_2$ system computed by MQCT and plotted versus three major characteristics of the energy transfer process: the change $\Delta E$ of the internal rotational energy of collision partners, the change $\Delta j_1$ of rotational angular momentum quantum number of $H_2O$, and the change $\Delta\tau_1$ of the pseudo-quantum number of $H_2O$, which is defined as $\tau_1 = k_a - k_c$ and varies between $-j_1$ and $+j_1$ for every value $j_1$ of water. These dependencies are plotted separately for four symmetries of $H_2O$ + $H_2$ system (100 initial rotational states of water are included in each case), for five different initial states of the projectile $H_2$, and for all possible state-to-state transitions in $H_2$. In Figs. 5, 6 and 7 of the main text these data are presented only for *p*-$H_2O$ + *p*-$H_2$ at collision energy $U$ ~ 704 cm$^{-1}$. All other symmetries of $H_2O$ + $H_2$ and one more value of collision energy $U$ = 12000 cm$^{-1}$ are analyzed in Figs. S5-S25 of Supplemental Information. To the best of our knowledge this information has never been presented before. It permits us to understand the transfer of energy between the rotational states of two collisional partners, and from the internal rotational states of the molecule-molecule system (as a whole) to the relative translational motion of two collisional partners.

In Fig. 5 we present the correlation between state-to-state transition cross sections $\sigma_{n'\to n''}$ and the rotational energy transfer $\Delta E$ in the range $-3000 < \Delta E <$ +3000 cm$^{-1}$ for *p*-$H_2O$ + *p*-$H_2$ system. Overall, 0.3 million transitions in $H_2O$ + $H_2$ are included in Fig. 5. The upper row of frames corresponds to *excitation* of $H_2$ starting from various initial states $j_2'$, while the bottom row corresponds to *quenching* of $H_2$. Different colors are used for transitions with different values of $\Delta j_2$ in $H_2$, namely $\Delta j_2 = \pm 2, \pm 4, \pm 6, \pm 8$ are shown by orange, grey, yellow, and maroon symbols, respectively. Transitions with elastic $H_2$ ($\Delta j_2 = 0$) are shown in blue in all the plots.

Qualitatively, the dependencies presented in Fig. 5 (and other figures for collision energy $U$ ~ 704 cm$^{-1}$, see SI) indicate a single-exponential behavior of collisional energy transfer. For transitions where the projectile $H_2$ remains *elastic* ($\Delta j_2 = 0$, blue datapoints) the energy transfer profile remains nearly symmetric around the elastic peak at $\Delta E = 0$, with quenching "wing" ($\Delta E < 0$) looking like a reflection of the excitation "wing" ($\Delta E > 0$) through at least three orders of magnitudes of cross section values. Only for very large energy transfer ($\Delta E \sim \pm 3000$ cm$^{-1}$) that correspond to very small cross sections, the deviations from this symmetry can be noticed, with excitation wing being somewhat higher compared to quenching. Overall, both wings are continuous and have a relatively well-defined upper boundary. The "peak" of blue dataset ($\Delta j_2 = 0$) remains near $\Delta E = 0$ through all the frames of Fig. 5 and in all relevant figures of Supplemental Information (see Figs. S5, S8, S11, S14, S17, S20, S23).

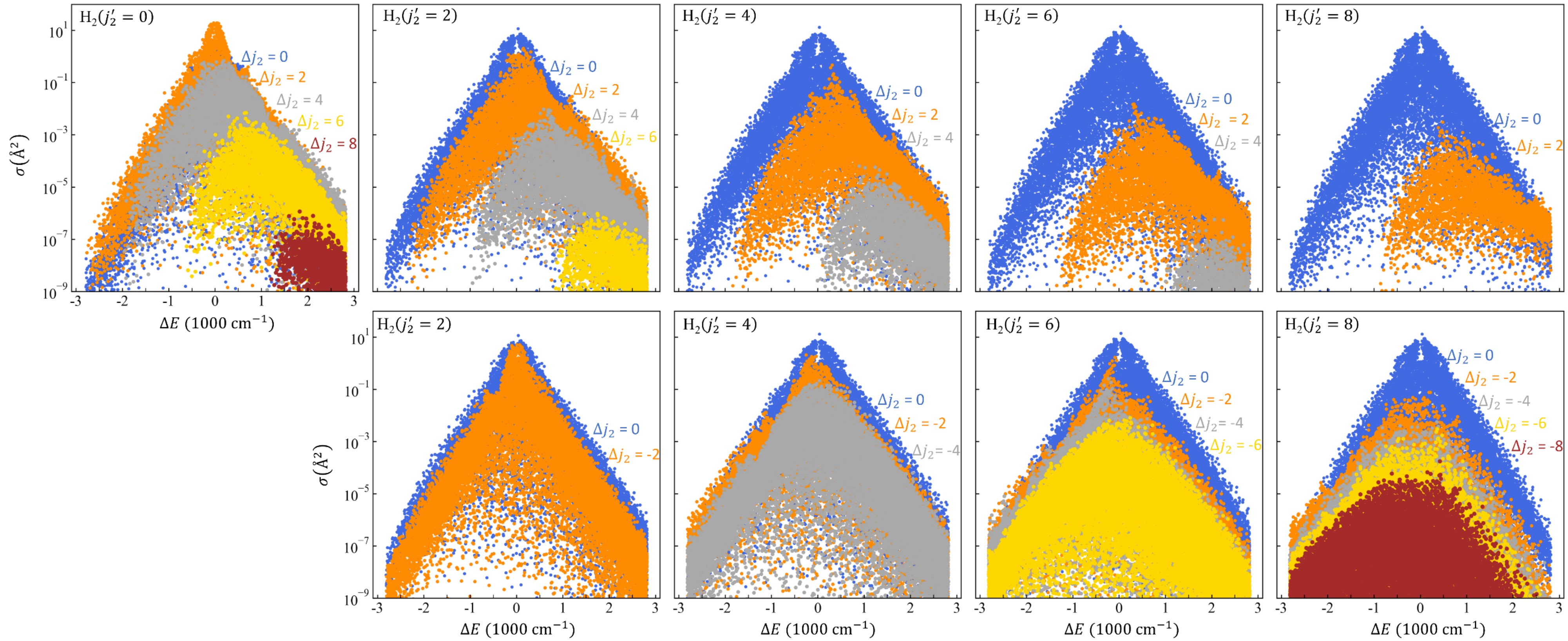

**Figure 5:** Correlation between the values of individual state-to-state transition cross sections $\sigma_{n' \to n''}$ and the overall transfer of internal rotational energy $\Delta E$ in *p*-$H_2O$ + *p*-$H_2$ system at collision energy $U$ ~ 704 $cm^{-1}$. Here 100 initial states of water are considered. The initial state of $H_2$ molecule is given in the upper left corner of each frame. Top and bottom rows of frames correspond to excitation and quenching of $H_2$. Colors represent different transitions in $H_2$: $\Delta j_2 = 0, \pm 2, \pm 4, \pm 6, \pm 8$ are shown by blue, orange, grey, yellow, and maroon symbols, respectively.

However, the state-to-state transitions with $\Delta j_2 \neq 0$ in which the energy transfer process not only involves the states of target $H_2O$ molecule, but also extends onto the states of projectile $H_2$ (all colors other than blue in Fig. 5) behave differently. As we examine the progression of frames going from left to right through Fig. 5, the following trend becomes apparent: the values of cross sections progressively drop as the energy transfer to/from the projectile increases (measured by the extend of *inelasticity* $\Delta j_2 \neq 0$ of $H_2$) and, simultaneously, the datapoints move significantly towards more positive values of $\Delta E$ when $H_2$ undergoes excitation ($\Delta j_2 > 0$, upper row of frames in Fig. 1), and they move (although less) towards more negative values of $\Delta E$ when $H_2$ undergoes quenching ($\Delta j_2 < 0$, lower row of frames in Fig. 1). This behavior means that the energy transfer processes to/from $H_2$, and to/from $H_2O$ happen differently. Consider the upper leftmost frame of Fig. 5. If $H_2$ would only absorb the rotational energy released by excited $H_2O$, then the datasets of all colors would remain centered around the total $\Delta E = 0$, but this does not happen. The fact that gray, yellow and maroon datapoints for the excitations of $H_2$ with $\Delta j_2$ = 4, 6 and 8 move towards large positive $\Delta E$ means that the rotational excitation of $H_2$ solely by the transfer of rotational energy from $H_2O$ is less likely than a process in which $H_2$ takes the rotational energy released by $H_2O$, plus some energy of the relative translational motion of collision partners. Now consider the lower rightmost frame of Fig. 5. Again, if $H_2O$ would be able to absorb efficiently all rotational energy released by $H_2$, then the datasets of all colors would remain centered around the $\Delta E = 0$, but this does not happen either. The fact that gray, yellow and maroon datapoints for the quenching of $H_2$ with $\Delta j_2$ = −4, −6 and −8 move towards negative $\Delta E$ means that the rotational quenching of $H_2$ solely by the transfer of its rotational energy to $H_2O$ is less likely than a process in which $H_2$ releases some rotational energy to $H_2O$, but some energy goes to the relative translational motion of collision partners. The fact that these shifts off-center are smaller in the lower frames of Fig. 5 than in its upper frames (for yellow and maroon datasets in particular), means that the transfer of energy from the rotationally excited states of $H_2$ to the rotational states of $H_2O$ is more direct (includes less rotational-to-translational energy transfer) compared to the reverse process. Same trends are seen in the relevant figures in Supplemental Information: Fig. S5, S8, S11, S14, S17, S20, and S23.

It is noteworthy that all transitions in which the projectile $H_2$ remains elastic ($\Delta j_2 = 0$, blue datapoints) maintain their prominence as one of the largest peaks in all frames of Fig. 5. One exception is the upper leftmost frame of Fig. 5 for $H_2O$ + $H_2(j_2' = 0)$, where transitions with $\Delta j_2 =$ 2 (orange) exhibit cross sections $\sigma_{n'\to n''}$ comparable to those with $\Delta j_2 = 0$. Besides this one exception, cross sections for all other inelastic transitions (all colors except blue) are smaller than the elastic once. For transitions with given initial $j_2'$ of $H_2$, the values of inelastic cross sections decrease, often by an order of magnitude or more, when the value of $H_2$ inelasticity $\Delta j_2$ is raised (compare different colors within given frame in Fig. 5) and this effect is more pronounced for transitions where $H_2$ is excited ($\Delta j_2 > 0$, upper frame) than for transitions where $H_2$ is quenched ($\Delta j_2 < 0$, lower frame in Fig. 5). Additionally, for transitions with given $\Delta j_2$ of $H_2$ the values of cross sections decrease as the initial excitation $j_2'$ of $H_2$ is raised. One can notice this trend by monitoring evolution of each color going from left to right through the frames of Fig. 5 (in particular, orange, gray and yellow datasets). Same trends are also observed in the relevant figures of Supplemental Information: Figs. S5, S8, S11, S14, S17, S20 and S23).

Importantly, the width of each wing in Fig. 5 (the spread of datapoints along the vertical direction) is rather large, namely: for a given value of collisional energy transfer $\Delta E$ the values of cross sections $\sigma_{n'\to n''}$ for individual state-to-state transitions vary withing two orders of magnitudes. This property is seen clearly for transitions with elastic $H_2$ (blue datasets in the upper row of frames in Fig. 5 for $j_2' = 4$, 6 and 8) but also for inelastic transitions in $H_2$ (orange and gray datasets in the lower row of frames in Fig. 5 for $j_2' = 2$ and 4). This feature suggests that one or several more factors, other than $\Delta E$, have a significant effect on the collisional energy transfer.

To identify these factors, we examined the dependencies of the state-to-state transition cross sections $\sigma_{n'\to n''}$ on the change of rotational quantum numbers of $H_2O$ molecule, $\Delta j_1$ and $\Delta \tau_1$. These are presented in Fig. 6 and 7, respectively, for *p*-$H_2O$ + *p*-$H_2$ at collision energy $U \sim 704$ cm$^{-1}$. It should be stressed that Figs. 6 and 7 contain the same exactly cross section data as Fig. 5, but the abscissa is different in all three figures, giving us three alternative raster images and exhibiting the dependence of cross sections on three variables: $\Delta E$, $\Delta j_1$ and $\Delta \tau_1$. Overall, the behavior of datapoints in Figs. 6 and 7 is similar to that of Fig. 5. All dependencies are centered around the elastic process with $\Delta j_1 = 0$ and $\Delta \tau_1 = 0$, are roughly symmetric, and exhibit trends close to a single-exponential energy transfer.

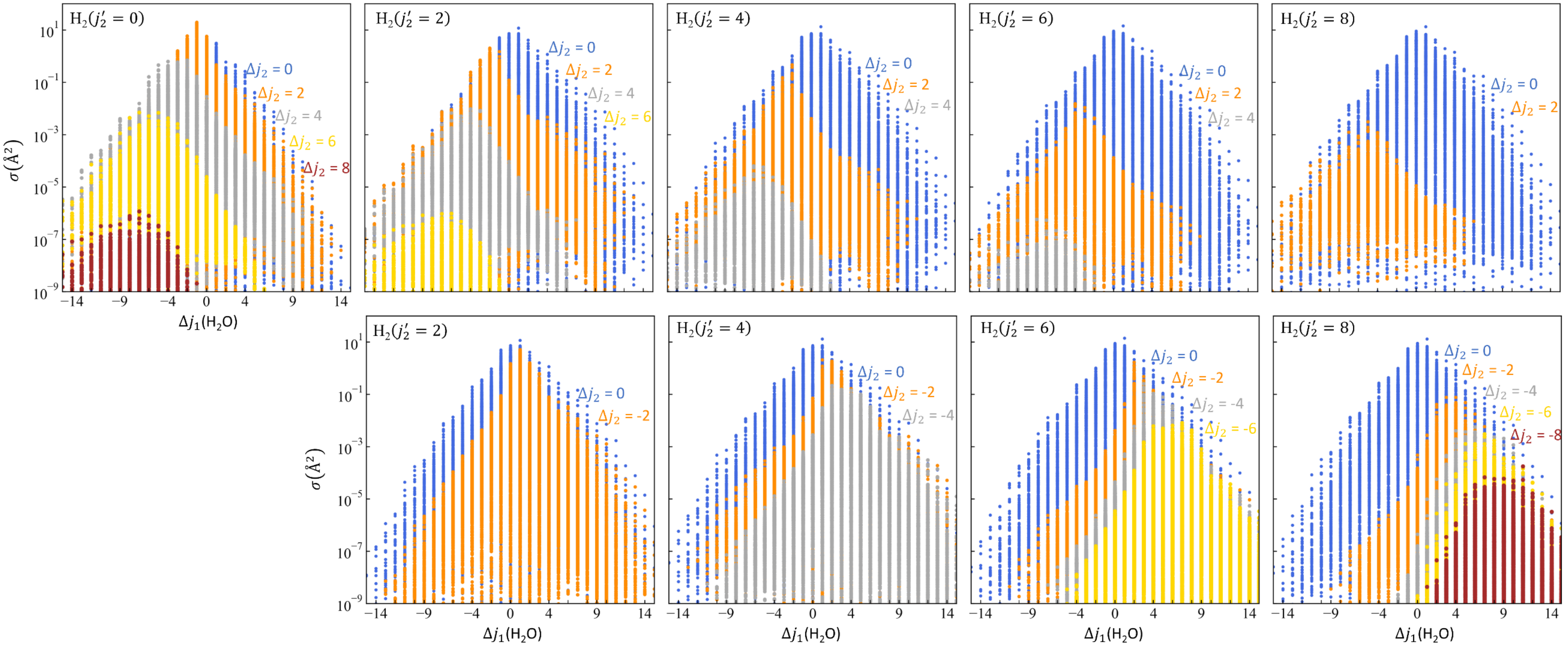

**Figure 6:** Correlation between the values of individual state-to-state transition cross sections $\sigma_{n' \to n''}$ and change in the rotational state of $H_2O$ ($\Delta j_1$) in the $p$-$H_2O$ + $p$-$H_2$ system, at $U$ ~ 704 $cm^{-1}$. Here 100 initial states of water are considered. The initial state of $H_2$ molecule is given in the upper left corner of each frame. Top and bottom rows of frames correspond to excitation and quenching of $H_2$. Colors represent different transitions in $H_2$: $\Delta j_2 = 0, \pm 2, \pm 4, \pm 6, \pm 8$ are shown by blue, orange, grey, yellow, and maroon symbols, respectively.

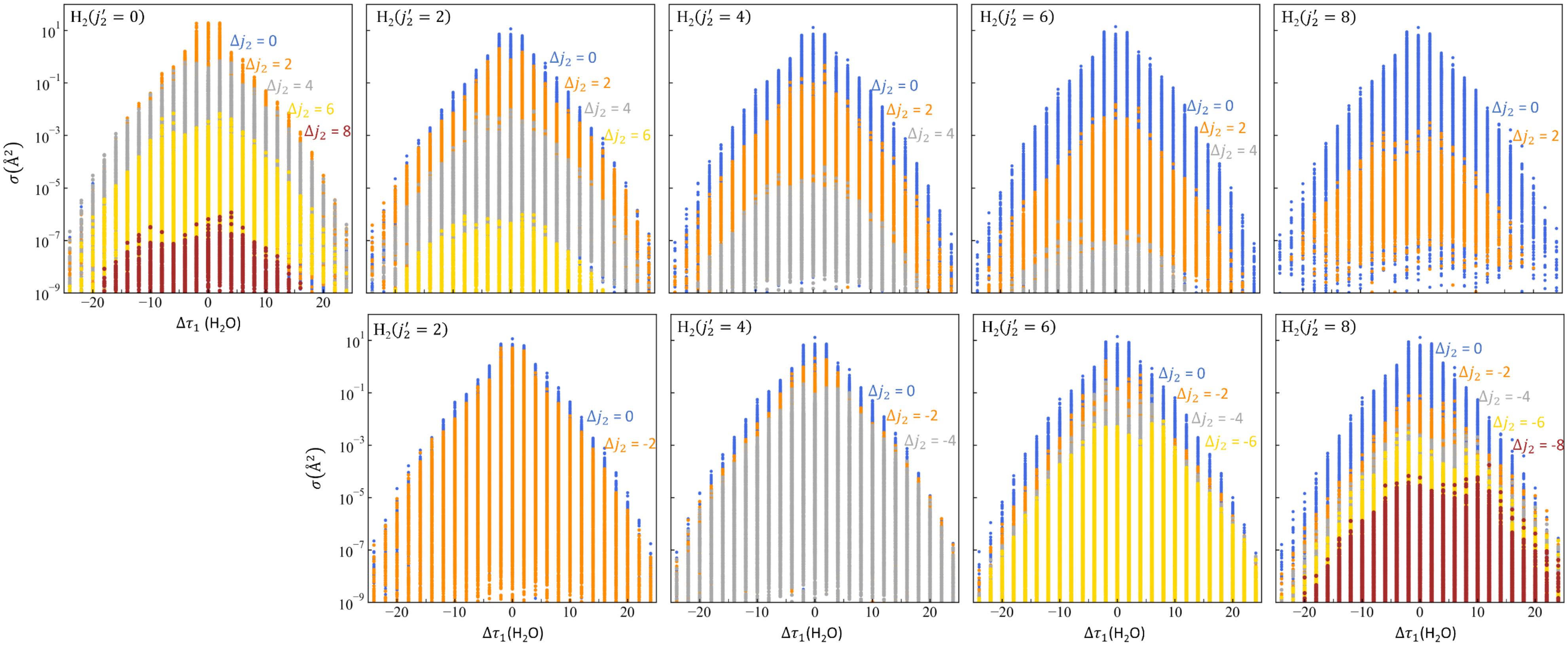

**Figure 7:** Correlation between the values of individual state-to-state transition cross sections $\sigma_{n'\to n''}$ and change in the $\tau_1$ of $H_2O$ ($\Delta\tau_1$) in the *p*-$H_2O$ + *p*-$H_2$ system, at $U$ ~ 704 cm$^{-1}$. Here 100 initial states of water are considered. The initial state of $H_2$ molecule is given in the upper left corner of each frame. Top and bottom rows of frames correspond to excitation and quenching of $H_2$. Colors represent different transitions in $H_2$: $\Delta j_2 = 0, \pm 2, \pm 4, \pm 6, \pm 8$ are shown by blue, orange, grey, yellow, and maroon symbols, respectively.

In Fig. 6, particularly if one looks at the upper leftmost and the lower rightmost frames, one can notice two very clear trends: as the magnitude of $\Delta j_2$ is raised the data points move towards large negative $\Delta j_1$ in the case when $H_2$ undergoes excitation ($\Delta j_2 > 0$), but they move towards large positive $\Delta j_1$ in the case when $H_2$ undergoes quenching ($\Delta j_2 < 0$). This behavior is opposite to what we saw in Fig. 5, and the reason for that is that in Fig. 6 the abscissa reflects the change of rotational angular momentum for water molecule only, while in Fig. 5 the abscissa gives the change of rotational energy of two collision partners together. (Note, if one tries to replot Fig. 5 using $\Delta E_1$ as abscissa, not shown here, one sees the trend similar to that of Fig. 6.) The explanation of this trend is also similar to that discussed above in the case of Fig. 5. In particular, if we look at the maroon dataset in Fig. 6, we see that for $\Delta j_2 = +8$ the datapoints shift towards $\Delta j_1 = -8$, while for $\Delta j_2 = -8$ the datapoints shift towards $\Delta j_1 = +8$, which reflects an efficient transfer of rotational angular momentum between the $H_2O$ molecule and the $H_2$ projectile.

If one compares Fig. 7 vs Figs. 5 and 6, one notices that the dependence of cross section $\sigma_{n' \to n''}$ on the change of pseudo-rotational quantum number of water $\Delta\tau_1$ is much more symmetric and stable. For all frames of Fig. 7, and all colors within each frame, the datapoints are centered around the origin $\Delta\tau_1 = 0$. Neither the initial state of the quencher $j_2'$, nor the amount of inelasticity of the quencher $\Delta j_2$ affect this property, which is in sharp contrast with Figs. 5 and 6.

Overall, the analysis of Figs. 6 and 7 clearly demonstrates that cross sections $\sigma_{n' \to n''}$ exhibit strong dependencies on $\Delta j_1$ and $\Delta\tau_1$ of $H_2O$. As $\Delta j_1$ changes by ±15, the values of cross sections drop by ~ 10 orders of magnitude on average. Similar, as $\Delta\tau_1$ changes by ±25, the values of cross sections change through the same 10 orders of magnitude, and this dependence is firm. Thus, we can conclude that both $\Delta j_1$ and $\Delta\tau_1$ need to be considered as parameters governing the collisional energy transfer, together with $\Delta E$. Therefore, analytic models of collisional energy transfer should, probably, employ expressions that take into account all these factors, for example:

$$p = \exp\left\{-\frac{\Delta E}{\varepsilon}\right\} \exp\left\{-\frac{\Delta j}{\gamma}\right\} \exp\left\{-\frac{\Delta\tau}{\theta}\right\}$$

Moreover, the results presented in Figs. 5-7 also indicate strong dependence of collisional energy transfer on inelasticity $\Delta j_2$ of the projectile $H_2$, and on the initial rotational state $j_2'$ of the projectile $H_2$, which makes the overall picture even more complex. Analysis of the data presented in the Supplemental Information shows similar trends for the other symmetries of *p/o*-$H_2O$ + *p/o*-$H_2$.

Of course, more complexity comes from the dependence of energy transfer on the translational energy of two collision partners. In Figs. S14 - 25 of Supplemental Information we present the same dependencies as in Figs. 5-7, but computed for very high collision energy, $U$ = 12000 cm$^{-1}$, and the readers are encouraged to inspect those results. Many trends seen in Figs. 5-7 remain relevant at high collision energy too, but there are several differences that we want to emphasize here. First of all, at high collision energy the amount of energy transfer is much larger, covering the range $-3000 < \Delta E <$ +8000 cm$^{-1}$ in Fig. S8. A significant part of this energy comes from the relative translational motion of collision partners and is sufficient to drive transitions in $H_2$ up to $\Delta j_2 = 10$ (green data points in Figs. S14 - 16). Importantly, the energy transfer profile is not symmetric anymore, with a much longer excitation wing that exhibits a double-exponential character. For transitions with elastic quencher ($\Delta j_2 = 0$) cross sections are slightly smaller at high collision energy, compared to those in Fig. 5, but for transitions with inelastic quencher ($\Delta j_2 \neq 0$) cross sections are larger than those in Fig. 5, and show somewhat less variation between excitation and quenching of the quencher (less difference between the upper and lower rows of frames in Fig. S14, S17, S20 and S23, compared to Fig. 5). Another feature that becomes very clear at high collision energy is a group of transitions with large cross sections and relatively small value of energy transfer, $\Delta E = \pm$500 cm$^{-1}$. They show up in Fig. S14, S16, S17, S19, S20, S22, S23 and S25 as a concentration of datapoints in the vicinity of elastic process at the top of each frame. In fact, these transitions are also present at low collision energy (e.g., in Figs. 5 and 7) but are less prominent in that case. These state-to-state transition processes correspond to $\Delta j_1 = 0, \pm 1$ and $\Delta \tau_1 = 0, \pm 1$. As $\Delta j_1$ and $\Delta \tau_1$ increase to $\pm 2$ and beyond, the values of cross sections drop sharply by almost an order of magnitude.

## VI. CONCLUSIONS

In this paper we presented a theory for the description of collisional energy transfer in a general asymmetric-top-rotor + linear rotor system using a mixed quantum/classical approach, in which the rotational states of collision partners are described quantum mechanically using time-dependent Schrodinger equation while the translational motion of scattering partners is described classically using a mean-field Ehrenfest trajectory method. This theory was applied to describe the energy transfer in $H_2O$ + $H_2$ system, for which the accurate full quantum results are also available from the time independent coupled-channel calculations (for lower rotational states of two collision partners) and can be used as a solid benchmark. We found a very good agreement between state-to-state transition rate coefficients computed by MQCT and the benchmark quantum data, for all four symmetries of the system (ortho- and para-water combined with ortho- and parahydrogen), for about 8000 individual state-to-state transitions, and through a broad range of temperature. The differences between the results of the two methods were shown to originate, largely, in different basis set sizes and in the Monte-Carlo sampling procedure employed to generate the initial conditions for MQCT trajectories. In principle, these sources of remaining errors can be eliminated if needed.

Massive parallelization of MQCT calculations permits us to run them efficiently with larger basis sets and at higher collision energies compared to what is typically feasible using the full quantum approach. Here we took MQCT calculations to the limit, by increasing the number of rotational states by more than a factor of two in both water target and hydrogen projectile, and increasing the maximum collision energy by 50%, relative to the previous calculations. The huge amount of data generated by MQCT was analyzed to obtain a broader picture of collisional energy transfer in molecule + molecule systems. The dependencies of inelastic cross sections (for individual state-to-state transitions) on the initial rotational state of the quencher $j_2$, on the extent of inelasticity of the quencher $\Delta j_2$, on the total rotational energy transfer in the molecule-molecule system $\Delta E$, and on the change of rotational quantum numbers of the target water molecule $\Delta j_1$ and $\Delta \tau_1$, all were plotted, analyzed, and discussed in detail. The findings permit us to better understand not only the $H_2O$ + $H_2$ system, but also the phenomenon of collisional energy transfer in general. State-to-state transition rate coefficients for $H_2O$ + $H_2$ system generated in this work will be deposited into the BASECOL database[61] for users in the astrophysical modelling community. In the future, these data can also be used to develop a simplified analytic model of collisional energy transfer in $H_2O$ + $H_2$.

**CONFLICTS OF INTEREST**

There are no conflicts to declare.

**AKNOWLEDGEMENTS**

This research was supported by NASA, grant number 80NSSC24K0208. D. Babikov acknowledges the support of Way Klingler Research Fellowship and of the Haberman-Pfletshinger Research Fund. CJ acknowledges the support of Schmitt Fellowship. D. Bostan acknowledges the support of Eisch Fellowship and the Bournique Memorial Fellowship. We used HPC resources at Marquette funded in part by the National Science foundation award CNS-1828649.

*Supplemental Information for:*

# "Mixed Quantum/Classical Theory (MQCT) Approach to the Dynamics of Molecule-Molecule Collisions in Complex Systems"

by Carolin Joy, Bikramaditya Mandal, Dulat Bostan, Marie-Lise Dubernet[2] and Dmitri Babikov[*]

*Chemistry Department, Marquette University, Milwaukee, Wisconsin 53201-1881, USA*

## I. DETAILED DESCRIPTION OF FULL QUANTUM CALCULATIONS

Quantum state-to-state rate coefficients (Daniel et al. 2011) are available for the rotational quenching of 45 levels of *o/p*-$H_2O$ by *o/p*-$H_2(j)$ for temperature ranging from T = 5-1500K, where the transitions among $H_2$ levels have been considered up to $j(H_2) = 4$ for some water transitions and for some temperatures. These rate coefficients were obtained using a 5D average of the 9D PES of Valiron et. al., 2008 and with quantum calculations for the dynamics of the nuclei. Daniel et al, 2011 calculations extended and completed the previous calculations of Dubernet et al, 2009 and of Daniel et al, 2010 on this system. A summary of the available transitions is given in Table S1, with the full set of calculated rate coefficients available in the BASECOL database (Dubernet et al., 2023).

**Table S1:** A summary of rotational quenching transitions available from full quantum calculations (Daniel et al, 2011). The entries are given in the form: N{ [ $j(H_2)$ transitions], [ ]} where N is the number of the target levels for which there is quenching (target levels are ordered by increasing energies). References are indicated by subscripts *a*, *b*, and *c*. An asterisk (*) marks transitions for which the rate coefficients are calculated between 1000-1500 K.

| Initial level of $H_2$ | *o*-$H_2O$ | *p*-$H_2O$ |
|---|---|---|
| $j_2 = 0$ | $45\{[0 \rightarrow 0], [0 \rightarrow 2]\}^a$ | $45\{[0 \rightarrow 0], [0 \rightarrow 2]\}$ |
| $j_2 = 2$ | $45\{[2 \rightarrow 0], [2 \rightarrow 2]\}^a; 10\{[2 \rightarrow 4]\}^{a,*}$ | $20\{[2 \rightarrow 0], [2 \rightarrow 2]\}$ |
| $j_2 = 4$ | $10\{[4 \rightarrow 2]^*, [4 \rightarrow 4]\}^a$ | none |
| $j_2 = 1$ | $45\{[1 \rightarrow 1]\}^c;\ 20\{[1 \rightarrow 3]\}^c$ | $45\{[1 \rightarrow 1]\}^{b,c};\ 20\{[1 \rightarrow 3]\}^b$ |
| $j_2 = 3$ | $5\{[3 \rightarrow 1]\}^c;\ 5\{[3 \rightarrow 3]\}^c$ | $10\{[3 \rightarrow 1]\}^b;\ 10\{[3 \rightarrow 3]\}^b$ |

*a)* Dubernet et al, 2009
*b)* Daniel et al, 2010
*c)* Daniel et al, 2011

[2] Observatoire de Paris, PSL University, Sorbonne Université, CNRS, LERMA, Paris, France
[*] Author to whom all correspondence should be addressed; electronic mail: dmitri.babikov@mu.edu
Chemistry Department, Wehr Chemistry Building, Marquette University, Milwaukee, Wisconsin 53201-1881, USA

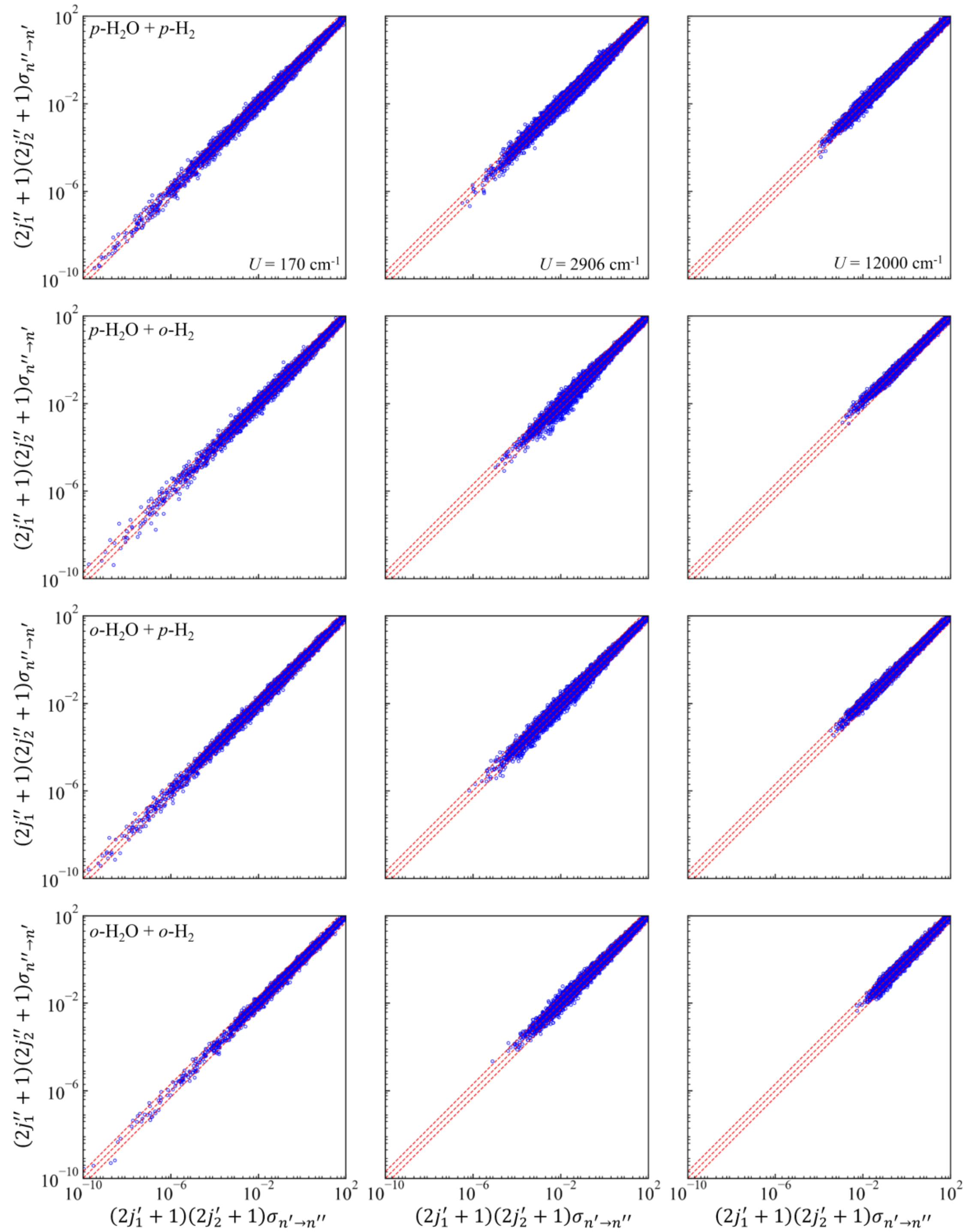

**Figure S1:** The comparison of MQCT state-to-state transition cross sections (in Å$^2$) for quenching and excitation directions of 118838 individual transitions in $H_2O$ + $H_2$ system at three values of collision energy $U$. The deviation of datapoints from the diagonal line indicates the departure from the principle of microscopic reversibility. The factor of 2 difference is shown by red dashed lines. Three columns correspond to three collision energies while four rows correspond to four symmetries of $H_2O$ + $H_2$ as indicated in the figure. In each case, 100 initial states of $o/p$-$H_2O$ are combined with two initial states of the projectile, either $j_2 = 0,\ 2, 4$ of $p$-$H_2$ or $j_2 = 1,\ 3$ of $o$-$H_2$.

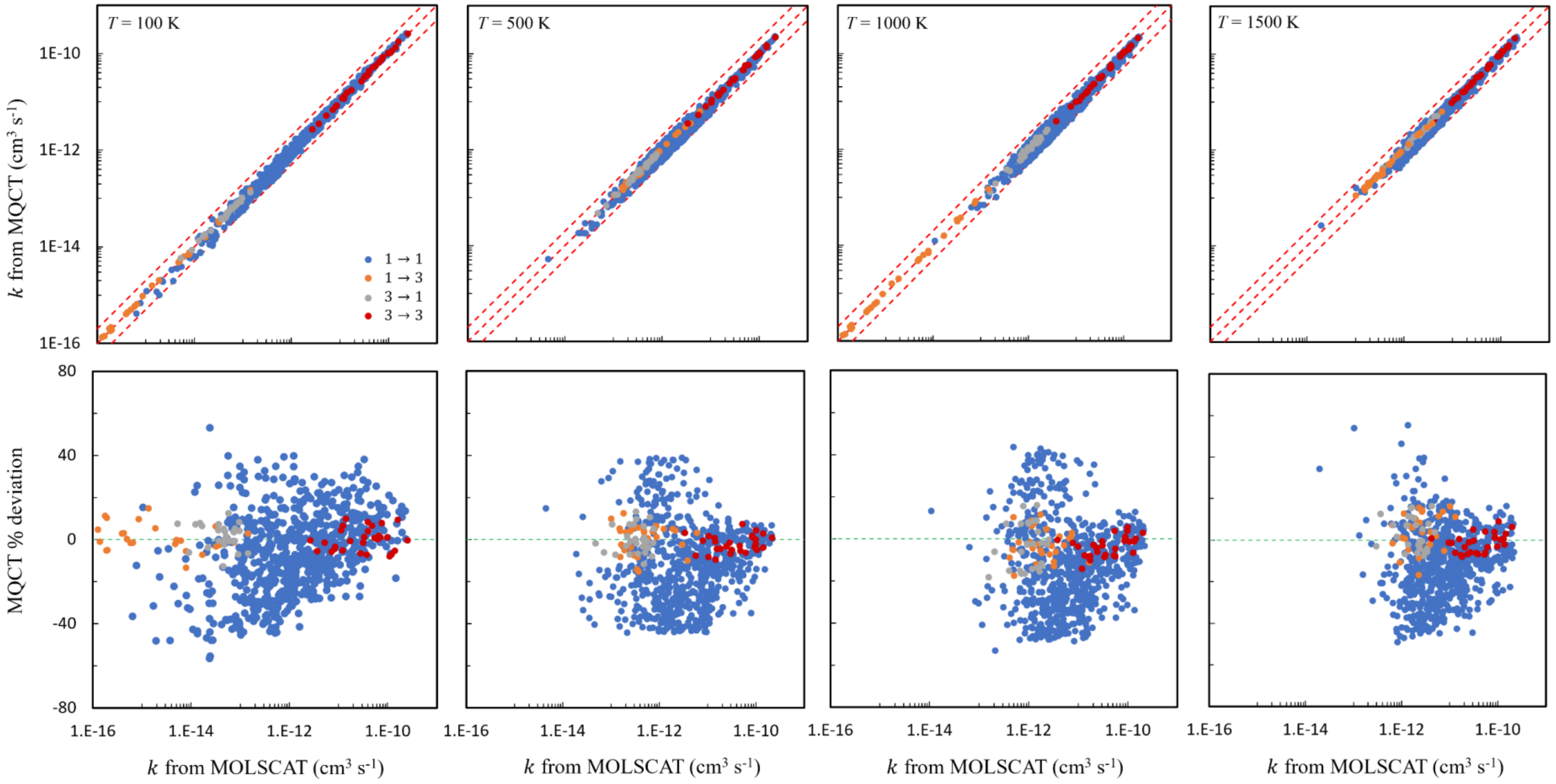

**Figure S2:** Comparison of 1270 state-to-state transition rate coefficients for *p*-$H_2O$ + *o*-$H_2$ collision computed using MQCT (this work) *vs* those predicted by full quantum MOLSCAT calculations (Daniel et al. 2011). Columns correspond to four values of temperature as indicated in the figure. The upper row of frames gives a one-to-one comparison of rate coefficients, while the lower row of frames presents percent deviations of MQCT data relative to MOLSCAT. Color is used to differentiate transitions in the projectile, namely, 1→1, 1→3, 3→1, and 3→3 transitions in $H_2$ are represented by blue, orange, grey and maroon, respectively. Red dashed lines in the upper row represent a factor of 2 difference.

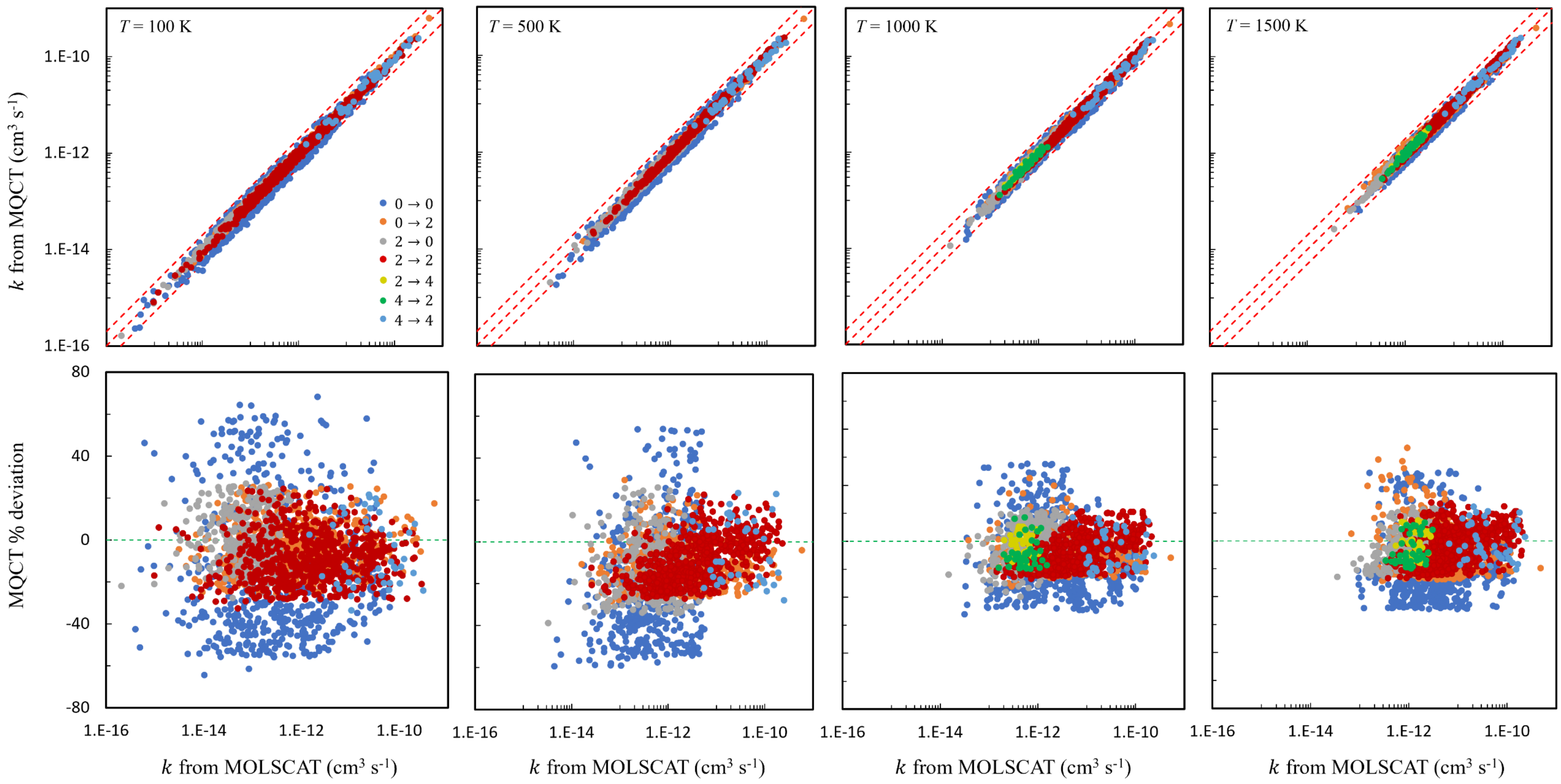

**Figure S3:** Comparison of 4095 state-to-state transition rate coefficients for *o*-$H_2O$ + *p*-$H_2$ collision computed using MQCT (this work) vs those predicted by full quantum MOLSCAT calculations (Daniel et al. 2011). Columns correspond to four values of temperature as indicated in the figure. The upper row of frames gives a one-to-one comparison of rate coefficients, while the lower row of frames presents percent deviations of MQCT data relative to MOLSCAT. Color is used to differentiate transitions in the projectile, namely, 0⟶0, 0⟶2, 2⟶0, 2⟶2, 2⟶4, 4⟶2 and 4⟶4 transitions in $H_2$ are represented by blue, orange, grey, maroon, yellow, green, and light blue respectively. Red dashed lines in the upper row represent a factor of 2 difference.

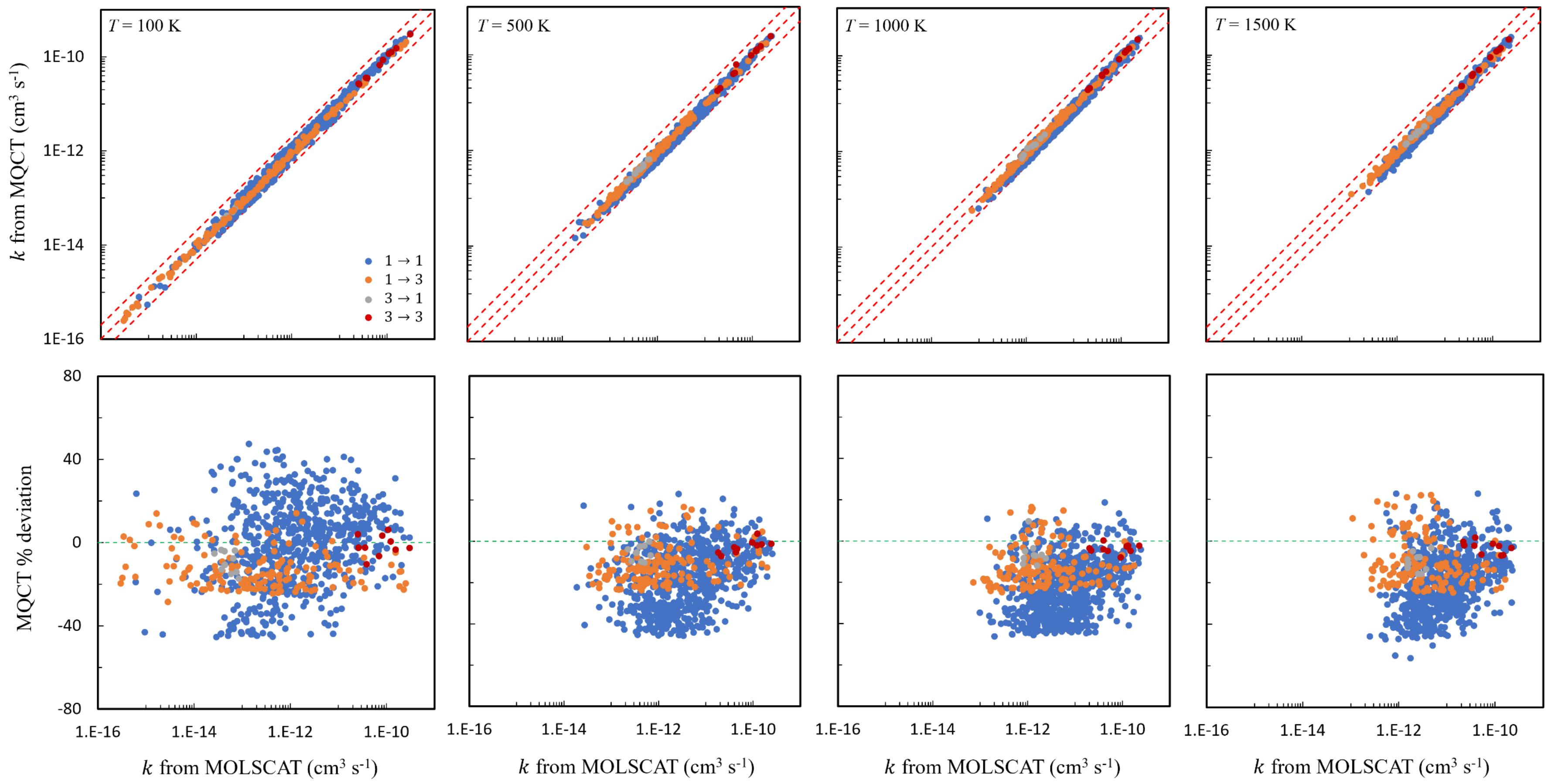

**Figure S4:** Comparison of 1200 state-to-state transition rate coefficients for *o*-$H_2O$ + *o*-$H_2$ collision computed using MQCT (this work) *vs* those predicted by full quantum MOLSCAT calculations (Daniel et al. 2011). Columns correspond to four values of temperature as indicated in the figure. The upper row of frames gives a one-to-one comparison of rate coefficients, while the lower row of frames presents percent deviations of MQCT data relative to MOLSCAT. Color is used to differentiate transitions in the projectile, namely, 1→1, 1→3, 3→1, and 3→3 transitions in $H_2$ are represented by blue, orange, grey and maroon, respectively. Red dashed lines in the upper row represent a factor of 2 difference.

**Table S2:** Average difference and RMS deviation (in %) for the comparisons shown in Fig. S2-S4 at four temperatures and for three symmetries of $H_2O$ + $H_2$, as indicated.

| $p$-$H_2O$ + $o$-$H_2$ | | | |
|---|---|---|---|
| Transitions in $H_2$ | T (K) | Average difference (%) | RMS of % deviation |
| $1 \rightarrow 1$ | 100 | -15 | 26 |
| | 500 | -12 | 24 |
| | 1000 | -11 | 22 |
| | 1500 | -10 | 20 |
| $1 \rightarrow 3$ | 100 | 4 | 16 |
| | 500 | 3 | 11 |
| | 1000 | -2 | 8 |
| | 1500 | 3 | 9 |
| $3 \rightarrow 1$ | 100 | 4 | 12 |
| | 500 | -2 | 7 |
| | 1000 | -4 | 10 |
| | 1500 | 3 | 8 |
| $3 \rightarrow 3$ | 100 | -0.1 | 6 |
| | 500 | -3 | 5 |
| | 1000 | -4 | 7 |
| | 1500 | -2 | 4 |

| $o$-$H_2O$ + $p$-$H_2$ | | | |
|---|---|---|---|
| Transitions in $H_2$ | T (K) | Average difference (%) | RMS of % deviation |
| $0 \rightarrow 0$ | 100 | -21 | 32 |
| | 500 | -18 | 30 |
| | 1000 | -17 | 28 |
| | 1500 | -15 | 26 |
| $0 \rightarrow 2$ | 100 | -10 | 14 |
| | 500 | -12 | 16 |
| | 1000 | -11 | 15 |
| | 1500 | -7 | 17 |
| $2 \rightarrow 0$ | 100 | 3 | 15 |
| | 500 | -14 | 20 |
| | 1000 | -0.8 | 13 |
| | 1500 | -8 | 15 |
| $2 \rightarrow 2$ | 100 | -8 | 15 |
| | 500 | -10 | 14 |
| | 1000 | -9 | 13 |
| | 1500 | -4 | 12 |
| $2 \rightarrow 4$ | 1000 | -3 | 10 |
| | 1500 | -5 | 9 |
| $4 \rightarrow 2$ | 1000 | -9 | 13 |
| | 1500 | -7 | 12 |
| $4 \rightarrow 4$ | 100 | -5 | 16 |
| | 500 | -4 | 15 |
| | 1000 | -6 | 14 |
| | 1500 | -1 | 13 |

| $o$-$H_2O$ + $o$-$H_2$ | | | |
|---|---|---|---|
| Transitions in $H_2$ | T (K) | Average difference (%) | RMS of % deviation |
| $1 \rightarrow 1$ | 100 | -5 | 24 |
| | 500 | -28 | 27 |
| | 1000 | -33 | 25 |
| | 1500 | -25 | 22 |
| $1 \rightarrow 3$ | 100 | -13 | 16 |
| | 500 | -11 | 14 |
| | 1000 | -12 | 15 |
| | 1500 | -10 | 16 |
| $3 \rightarrow 1$ | 100 | -10 | 11 |
| | 500 | -6 | 7 |
| | 1000 | -7 | 9 |
| | 1500 | -10 | 11 |
| $3 \rightarrow 3$ | 100 | -2 | 5 |
| | 500 | -4 | 4 |
| | 1000 | -5 | 6 |
| | 1500 | -4 | 15 |

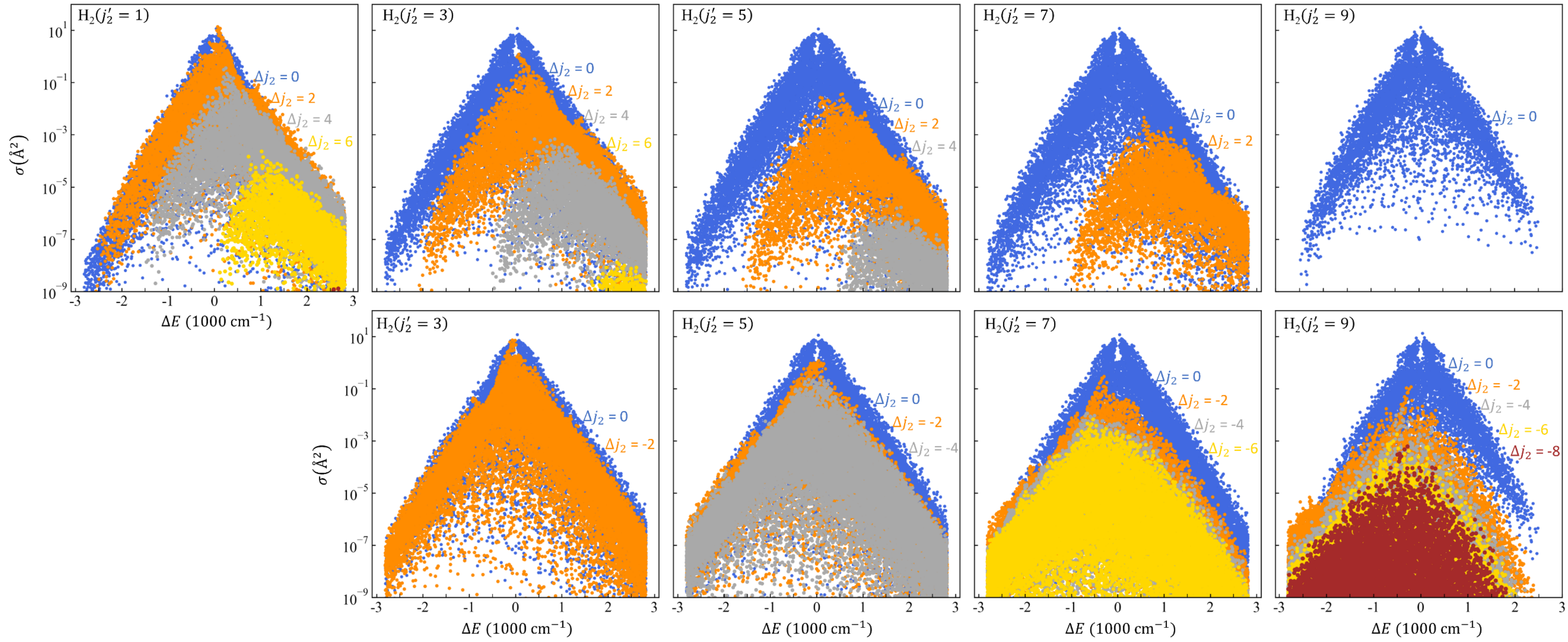

**Figure S5:** Correlation between the values of individual state-to-state transition cross sections $\sigma_{n' \to n''}$ and the overall transfer of internal rotational energy, $\Delta E$ in the *p*-$H_2O$ + *o*-$H_2$ system at collision energy $U \sim 704$ cm$^{-1}$. Here 100 initial states of water are considered. The initial state of $H_2$ molecule is given in the upper left corner of each frame. Top and bottom rows of frames correspond to excitation and quenching of $H_2$. Colors represent different transitions in $H_2$: $\Delta j_2 = 0, \pm 2, \pm 4, \pm 6, \pm 8$ are shown by blue, orange, grey, yellow, and maroon symbols, respectively.

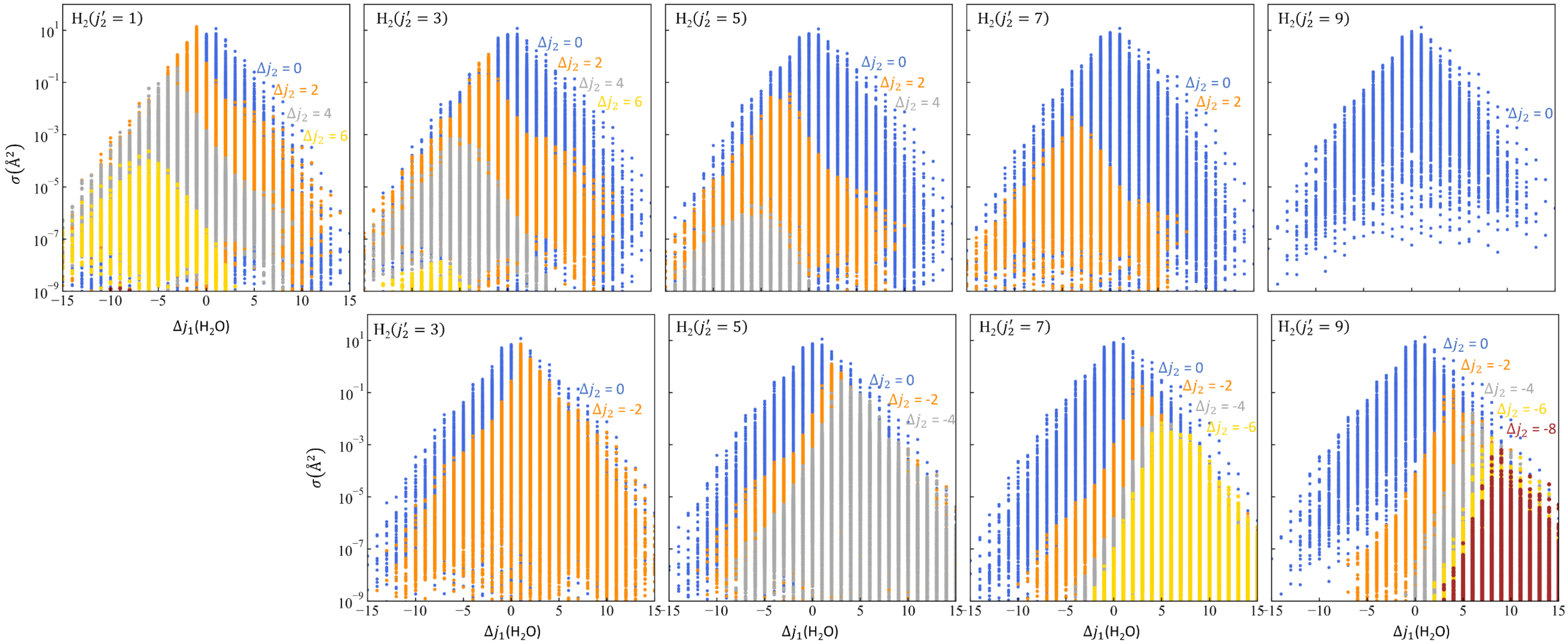

**Figure S6:** Correlation between the values of individual state-to-state transition cross sections $\sigma_{n'\rightarrow n''}$ and change in the rotational state of $H_2O$ ($\Delta j_1$) in the *p*-$H_2O$ + *o*-$H_2$ system, at $U$ ~ 704 $cm^{-1}$. Here 100 initial states of water are considered. The initial state of $H_2$ molecule is given in the upper left corner of each frame. Top and bottom rows of frames correspond to excitation and quenching of $H_2$. Colors represent different transitions in $H_2$: $\Delta j_2 = 0, \pm 2, \pm 4, \pm 6, \pm 8$ are shown by blue, orange, grey, yellow, and maroon symbols, respectively.

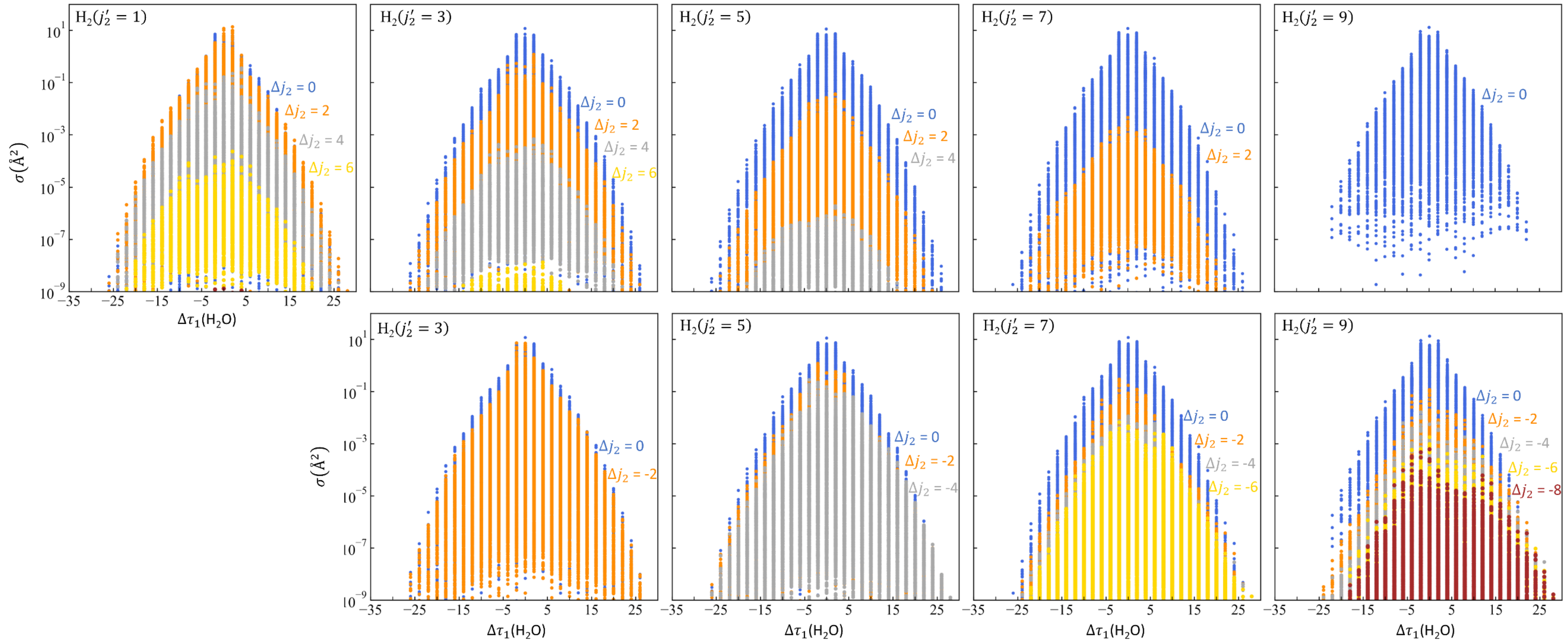

**Figure S7:** Correlation between the values of individual state-to-state cross sections $\sigma_{n' \to n''}$ and change in the $\tau_1$ of $H_2O$ ($\Delta\tau_1$) in the *p*-$H_2O$ + *o*-$H_2$ system, at $U \sim 704$ cm$^{-1}$. Here 100 initial states of water are considered. The initial state of $H_2$ molecule is given in the upper left corner of each frame. Top and bottom rows of frames correspond to excitation and quenching of $H_2$. Colors represent different transitions in $H_2$: $\Delta j_2 = 0, \pm 2, \pm 4, \pm 6, \pm 8$ are shown by blue, orange, grey, yellow, and maroon symbols, respectively.

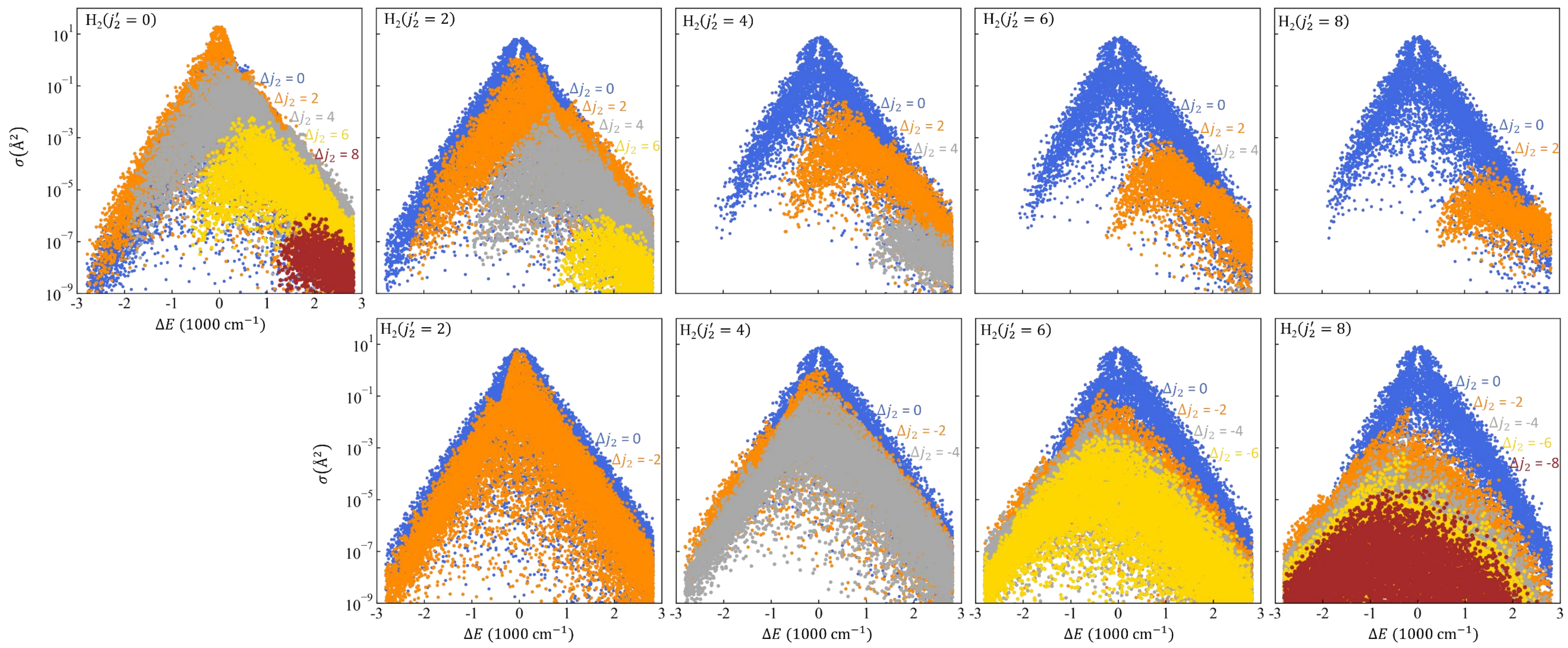

**Figure S8:** Correlation between the values of individual state-to-state transition cross sections $\sigma_{n' \to n''}$ and the overall transfer of internal rotational energy, $\Delta E$ in the *o*-$H_2O$ + *p*-$H_2$ system at collision energy $U$ ~ 704 cm$^{-1}$. Here 100 initial states of water are considered. The initial state of $H_2$ molecule is given in the upper left corner of each frame. Top and bottom rows of frames correspond to excitation and quenching of $H_2$. Colors represent different transitions in $H_2$: $\Delta j_2 = 0, \pm 2, \pm 4, \pm 6, \pm 8$ are shown by blue, orange, grey, yellow, and maroon symbols, respectively.

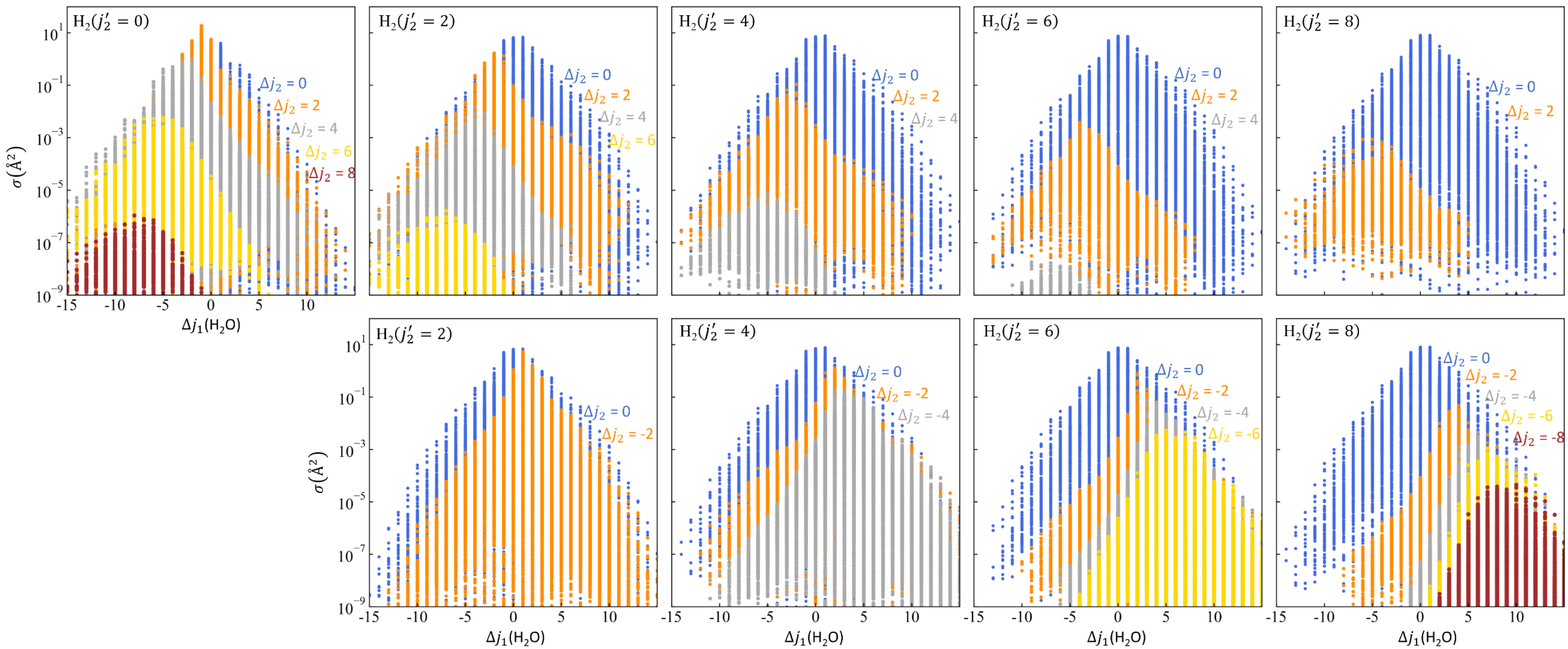

**Figure S9:** Correlation between the values of individual state-to-state transition cross sections $\sigma_{n'\to n''}$ and change in the rotational state of $H_2O$ ($\Delta j_1$) in the $o$-$H_2O$ + $p$-$H_2$ system, at $U$ ~ 704 cm$^{-1}$. Here 100 initial states of water are considered. The initial state of $H_2$ molecule is given in the upper left corner of each frame. Top and bottom rows of frames correspond to excitation and quenching of $H_2$. Colors represent different transitions in $H_2$: $\Delta j_2 = 0, \pm 2, \pm 4, \pm 6, \pm 8$ are shown by blue, orange, grey, yellow, and maroon symbols, respectively.

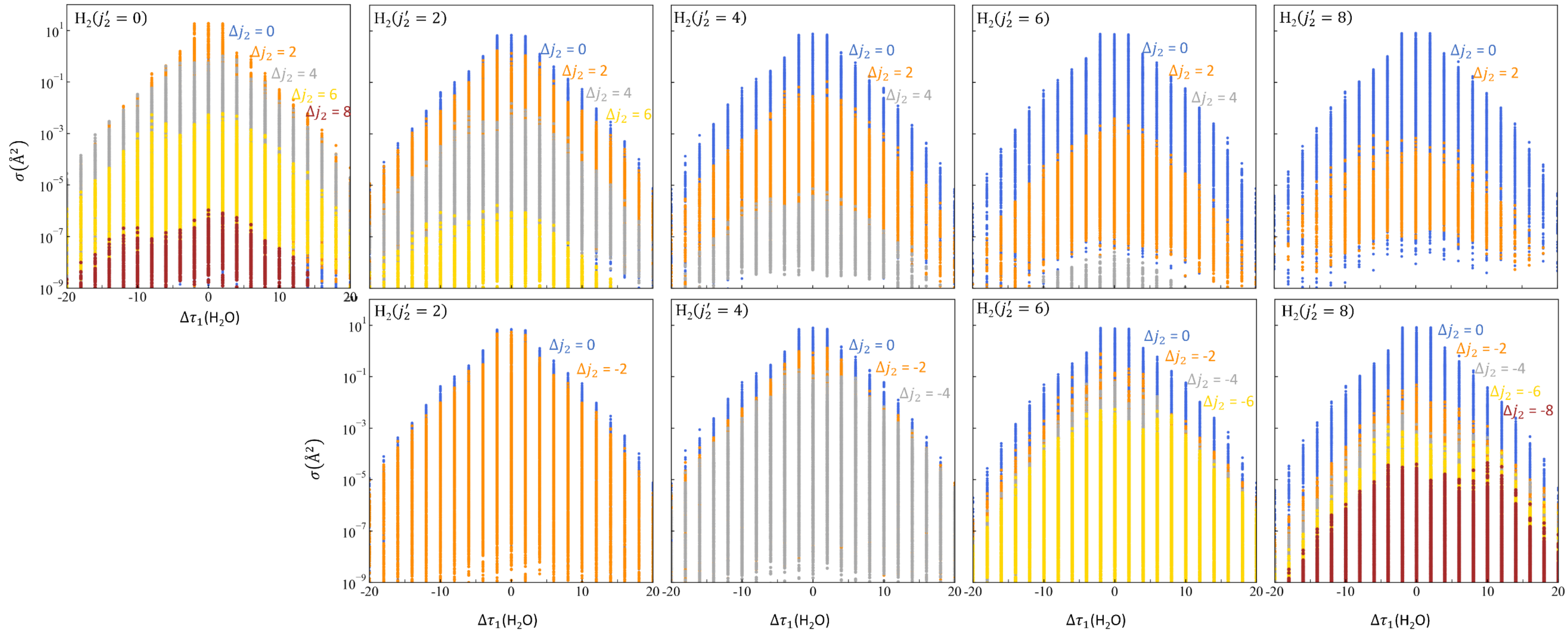

**Figure S10:** The correlation between the values of individual state-to-state transition cross sections $\sigma_{n'\to n''}$ and change in the $\tau_1$ of $H_2O$ ($\Delta\tau_1$) in the *o*-$H_2O$ + *p*-$H_2$ system, at $U \sim 704$ cm$^{-1}$. Here 100 initial states of water are considered. The initial state of $H_2$ molecule is given in the upper left corner of each frame. Top and bottom rows of frames correspond to excitation and quenching of $H_2$. Colors represent different transitions in $H_2$: $\Delta j_2 = 0, \pm2, \pm4, \pm6, \pm8$ are shown by blue, orange, grey, yellow, and maroon symbols, respectively.

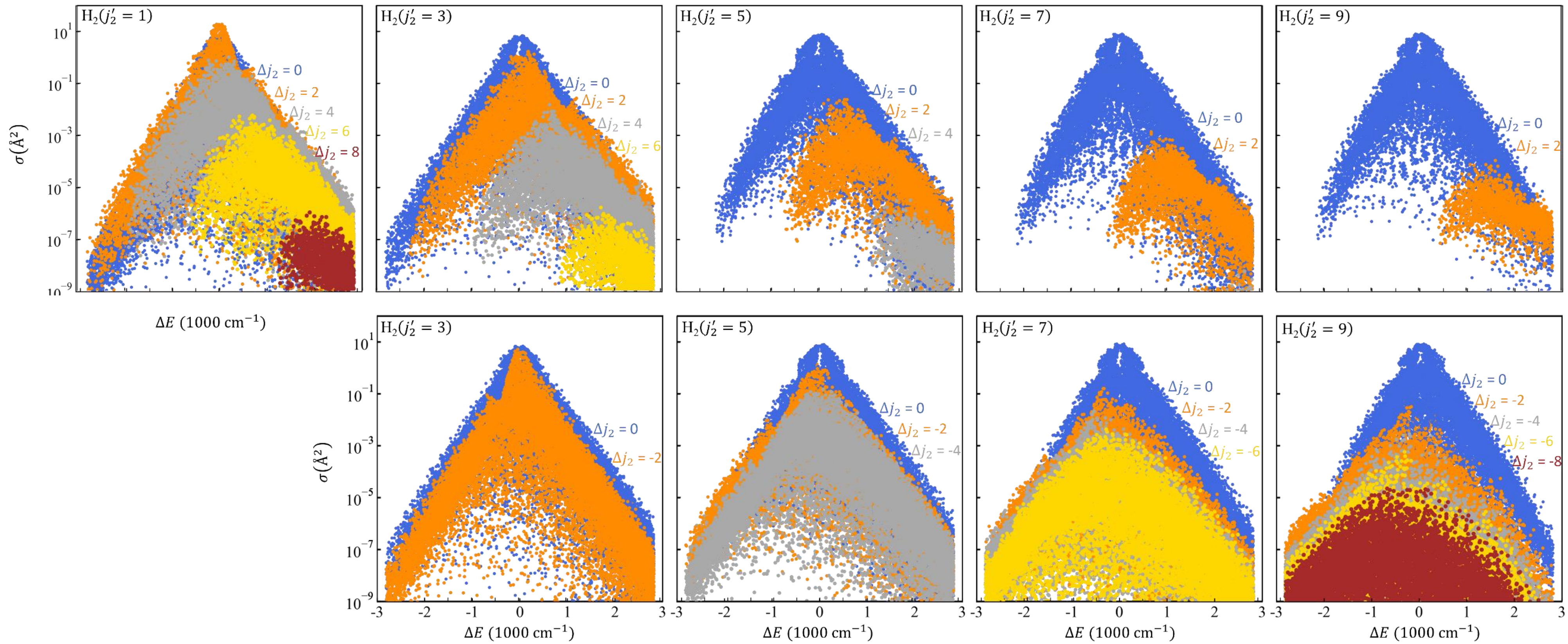

**Figure S11:** Correlation between the values of individual state-to-state transition cross sections $\sigma_{n' \to n''}$ and the overall transfer of internal rotational energy, $\Delta E$ in the *o*-$H_2O$ + *o*-$H_2$ system at collision energy $U$ ~ 704 cm$^{-1}$ Here 100 initial states of water are considered. The initial state of $H_2$ molecule is given in the upper left corner of each frame. Top and bottom rows of frames correspond to excitation and quenching of $H_2$. Colors represent different transitions in $H_2$: $\Delta j_2 = 0, \pm 2, \pm 4, \pm 6, \pm 8$ are shown by blue, orange, grey, yellow, and maroon symbols, respectively.

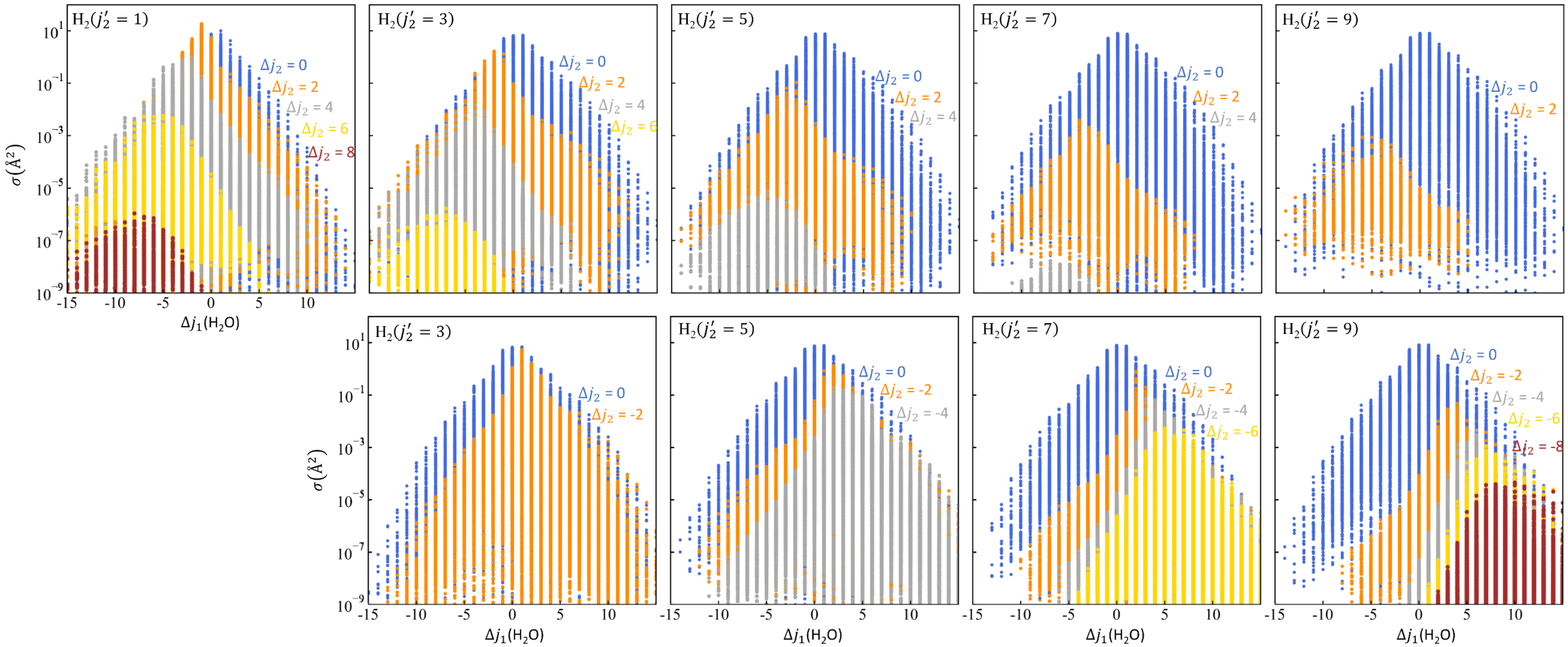

**Figure S12:** Correlation between the values of individual state-to-state transition cross sections $\sigma_{n' \to n''}$ and change in the rotational state of $H_2O$ ($\Delta j_1$) in the $o$-$H_2O$ + $o$-$H_2$ system, at $U \sim 704$ cm$^{-1}$. Here 100 initial states of water are considered. The initial state of $H_2$ molecule is given in the upper left corner of each frame. Top and bottom rows of frames correspond to excitation and quenching of $H_2$. Colors represent different transitions in $H_2$: $\Delta j_2 = 0, \pm 2, \pm 4, \pm 6, \pm 8$ are shown by blue, orange, grey, yellow, and maroon symbols, respectively.

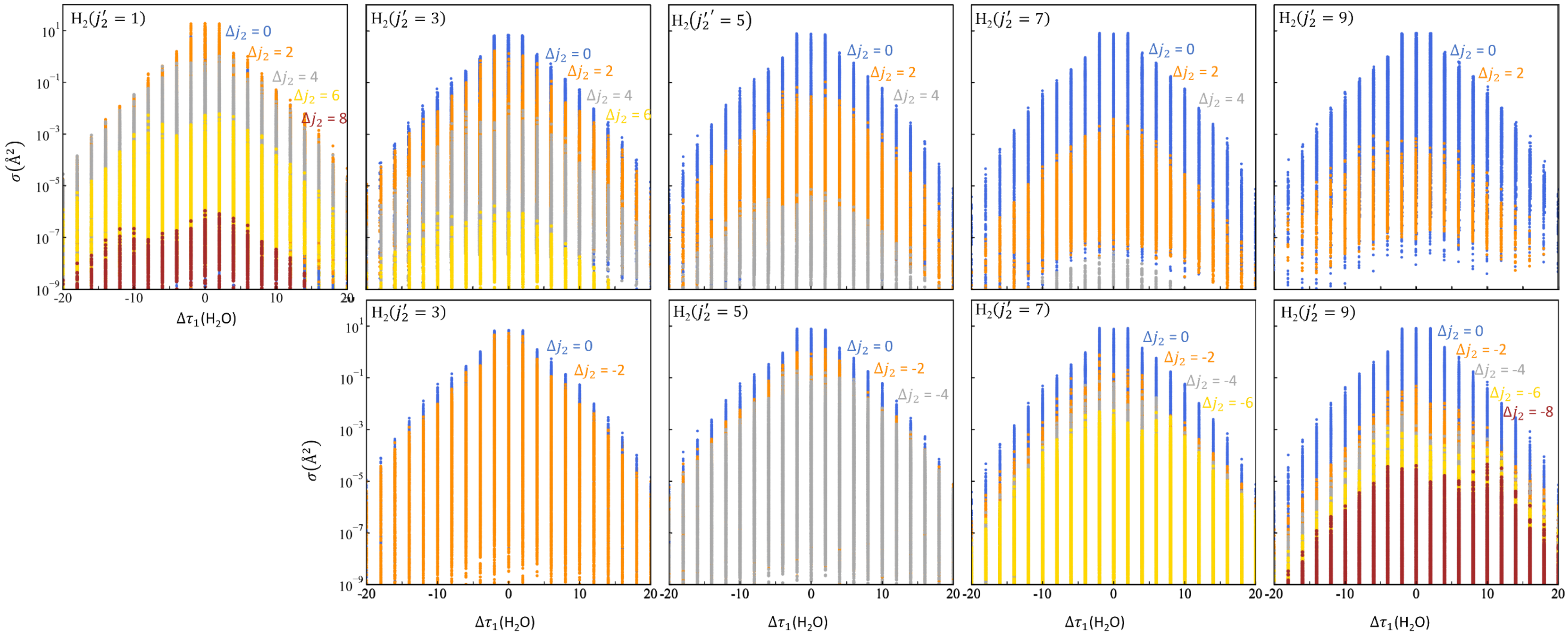

**Figure S13:** Correlation between the values of individual state-to-state transition cross sections $\sigma_{n' \to n''}$ and change in the $\tau_1$ of $H_2O$ ($\Delta\tau_1$) in the *o*-$H_2O$ + *o*-$H_2$ system, at $U \sim 704$ cm$^{-1}$. Here 100 initial states of water are considered. The initial state of $H_2$ molecule is given in the upper left corner of each frame. Top and bottom rows of frames correspond to excitation and quenching of $H_2$. Colors represent different transitions in $H_2$: $\Delta j_2 = 0, \pm 2, \pm 4, \pm 6, \pm 8$ are shown by blue, orange, grey, yellow, and maroon symbols, respectively.

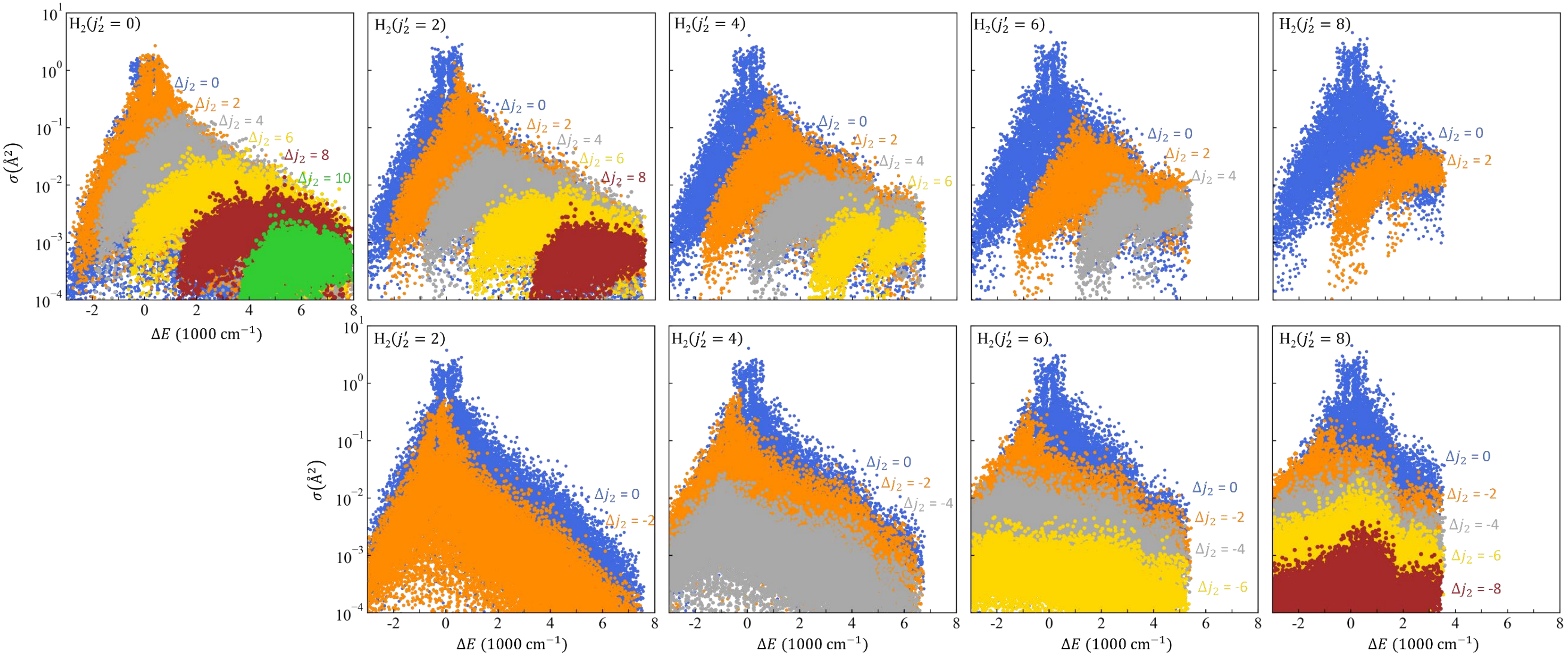

**Figure S14:** Correlation between the values of individual state-to-state transition cross sections $\sigma_{n' \to n''}$ and the overall extent of internal rotational energy $\Delta E$ in $p$-$H_2O$ + $p$-$H_2$ system at collision energy $U$ = 12000 cm$^{-1}$. Here 100 initial states of water are considered. The initial state of $H_2$ molecule is given in the upper left corner of each frame. Top and bottom rows of frames correspond to excitation and quenching of $H_2$. Colors represent different transitions in $H_2$: $\Delta j_2 = 0, \pm 2, \pm 4, \pm 6, \pm 8$ are shown by blue, orange, grey, yellow, and maroon symbols, respectively.

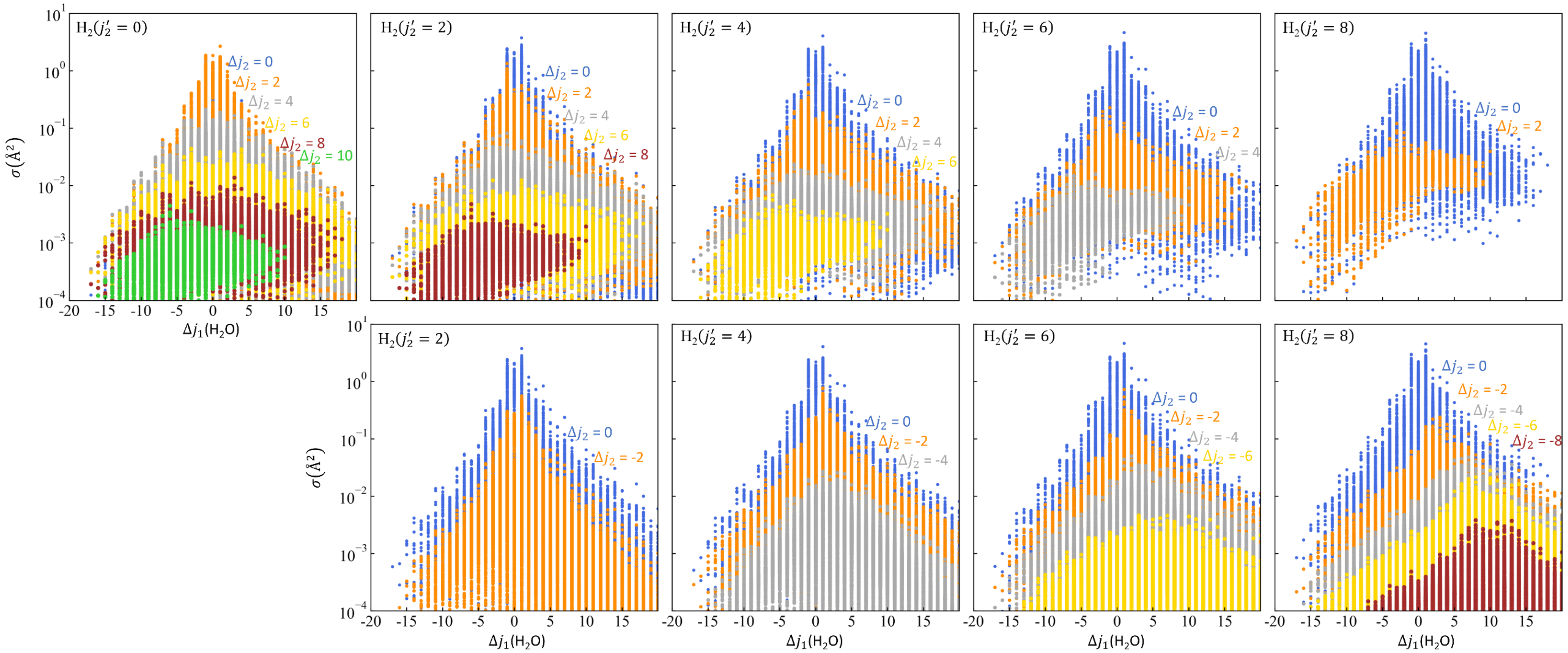

**Figure S15:** Correlation between the values of individual state-to-state transition cross sections $\sigma_{n' \to n''}$ and change in the rotational state of $H_2O$ ($\Delta j_1$) in the *p*-$H_2O$ + *p*-$H_2$ system, at $U$ = 12000 cm$^{-1}$. Here 100 initial states of water are considered. The initial state of $H_2$ molecule is given in the upper left corner of each frame. Top and bottom rows of frames correspond to excitation and quenching of $H_2$. Colors represent different transitions in $H_2$: $\Delta j_2 = 0, \pm 2, \pm 4, \pm 6, \pm 8$ are shown by blue, orange, grey, yellow, and maroon symbols, respectively.

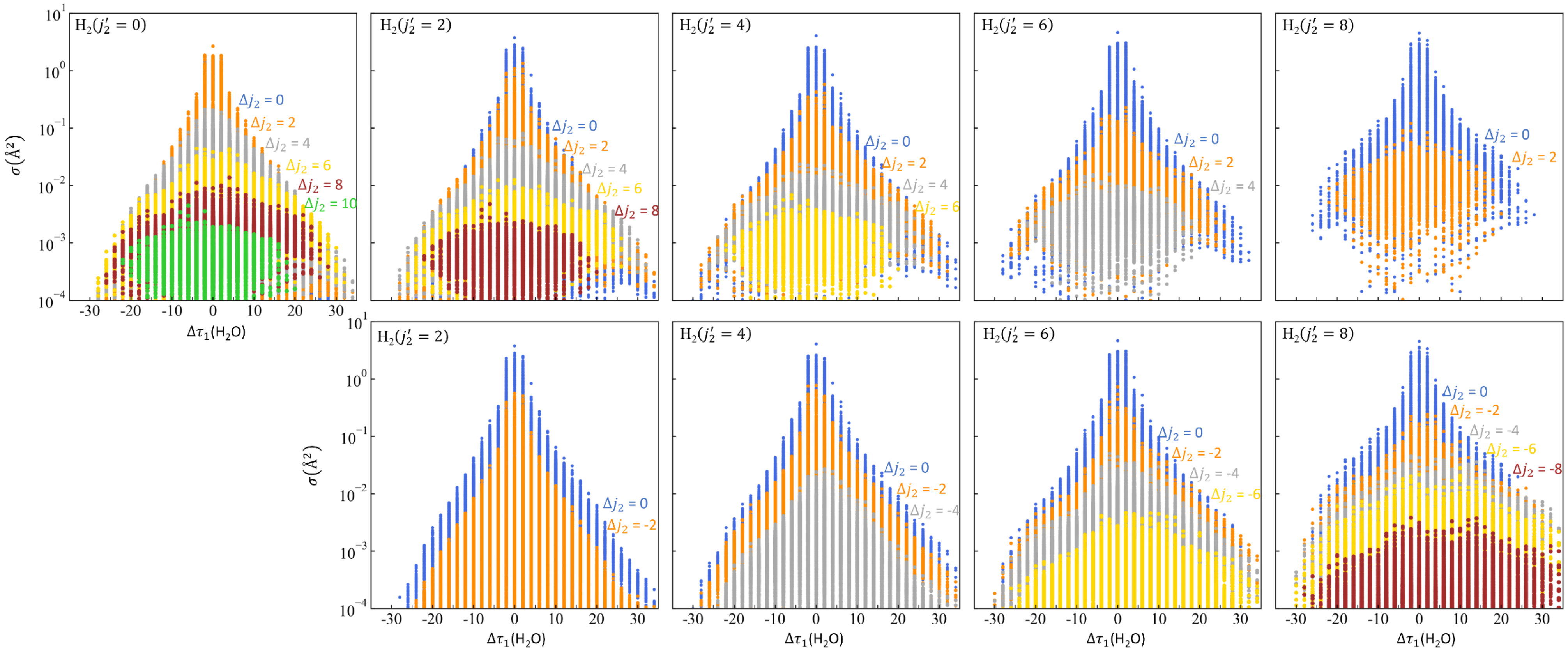

**Figure S16:** Correlation between the values of individual state-to-state transition cross sections $\sigma_{n'\to n''}$ and change in the $\tau_1$ of $H_2O$ ($\Delta\tau_1$) in the *p*-$H_2O$ + *p*-$H_2$ system, at *U* = 12000 cm$^{-1}$. Here 100 initial states of water are considered. The initial state of $H_2$ molecule is given in the upper left corner of each frame. Top and bottom rows of frames correspond to excitation and quenching of $H_2$. Colors represent different transitions in $H_2$: $\Delta j_2 = 0, \pm2, \pm4, \pm6, \pm8$ are shown by blue, orange, grey, yellow, and maroon symbols, respectively.

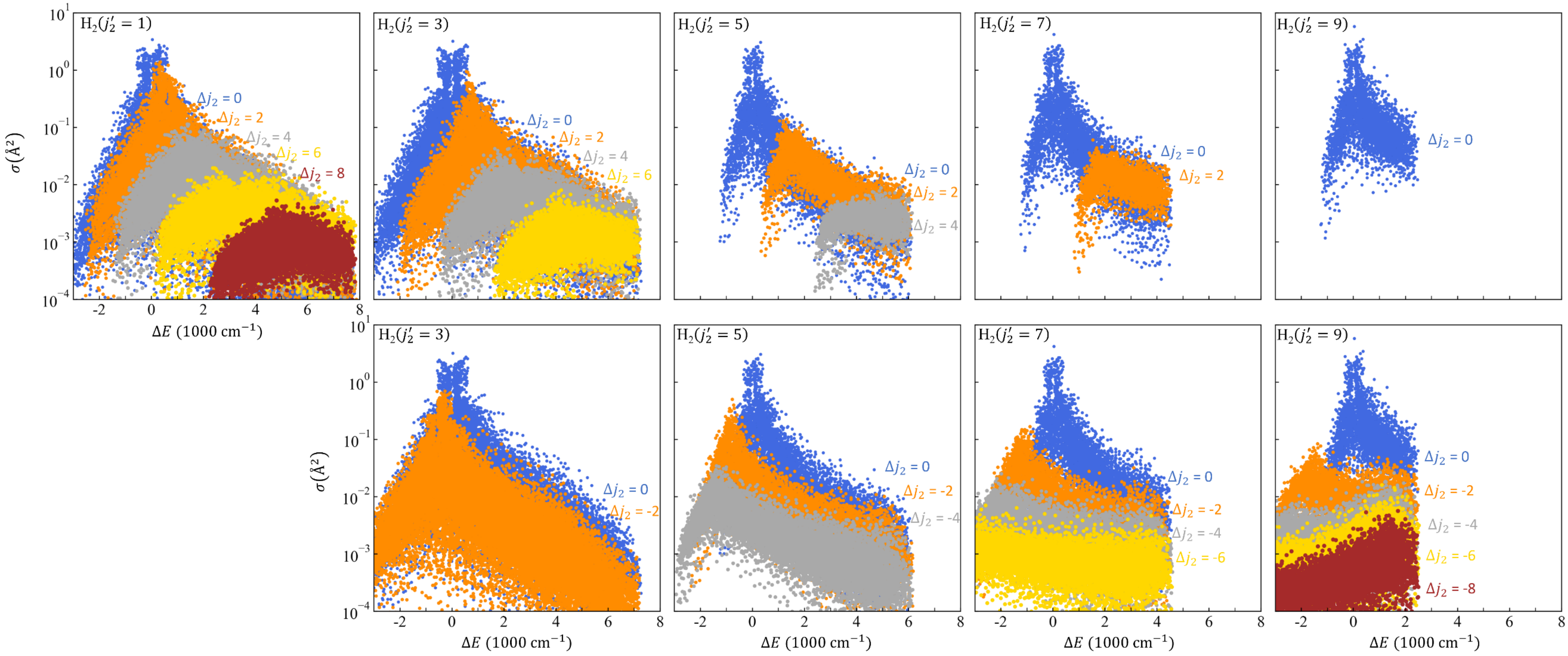

**Figure S17:** Correlation between the values of individual state-to-state transition cross sections $\sigma_{n' \to n''}$ and the overall extent of internal rotational energy $\Delta E$ in *p*-$H_2O$ + *o*-$H_2$ system at collision energy $U$ = 12000 cm$^{-1}$. Here 100 initial states of water are considered. The initial state of $H_2$ molecule is given in the upper left corner of each frame. Top and bottom rows of frames correspond to excitation and quenching of $H_2$. Colors represent different transitions in $H_2$: $\Delta j_2 = 0, \pm 2, \pm 4, \pm 6, \pm 8$ are shown by blue, orange, grey, yellow, and maroon symbols, respectively.

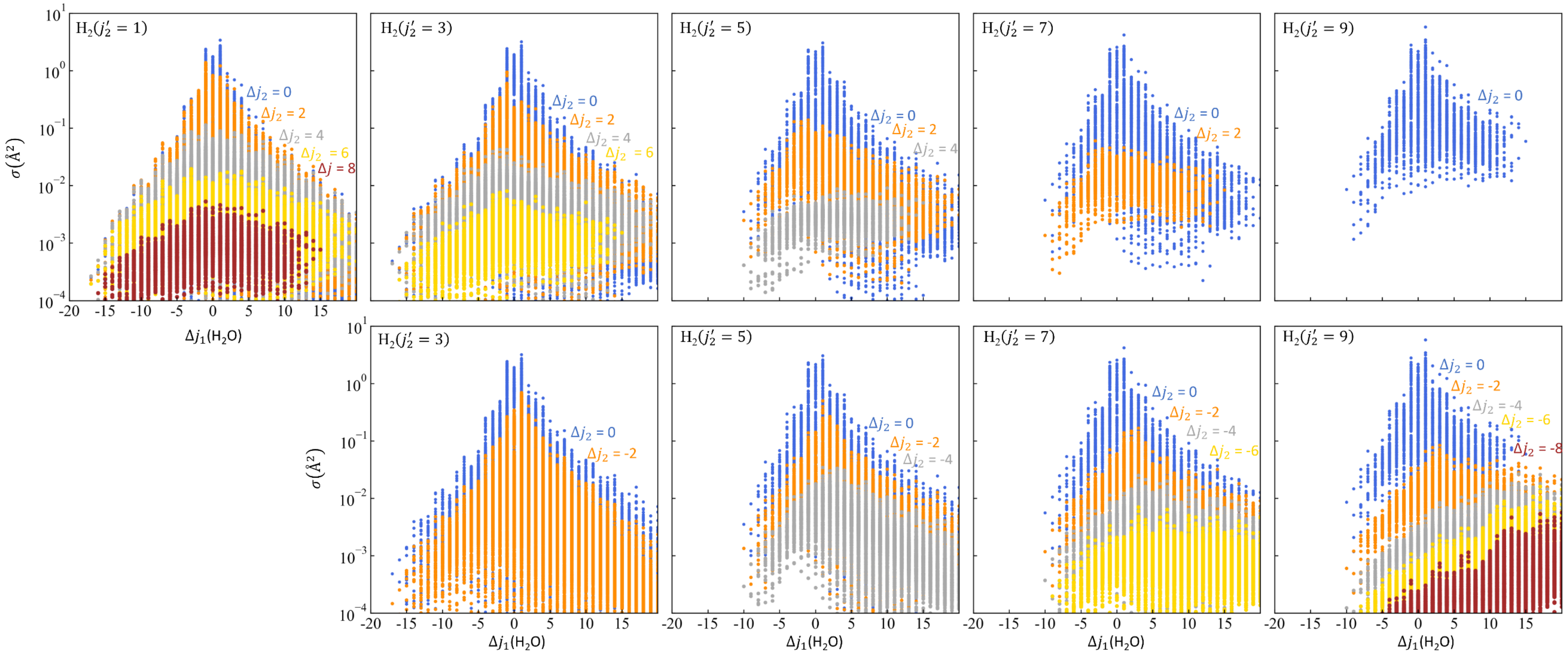

**Figure S18:** Correlation between the values of individual state-to-state cross sections $\sigma_{n'\to n''}$ and change in the rotational state of $H_2O$ ($\Delta j_1$) in the *p*-$H_2O$ + *o*-$H_2$ system, at $U$ = 12000 cm$^{-1}$. Here 100 initial states of water are considered. The initial state of $H_2$ molecule is given in the upper left corner of each frame. Top and bottom rows of frames correspond to excitation and quenching of $H_2$. Colors represent different transitions in $H_2$: $\Delta j_2 = 0, \pm 2, \pm 4, \pm 6, \pm 8$ are shown by blue, orange, grey, yellow, and maroon symbols, respectively.

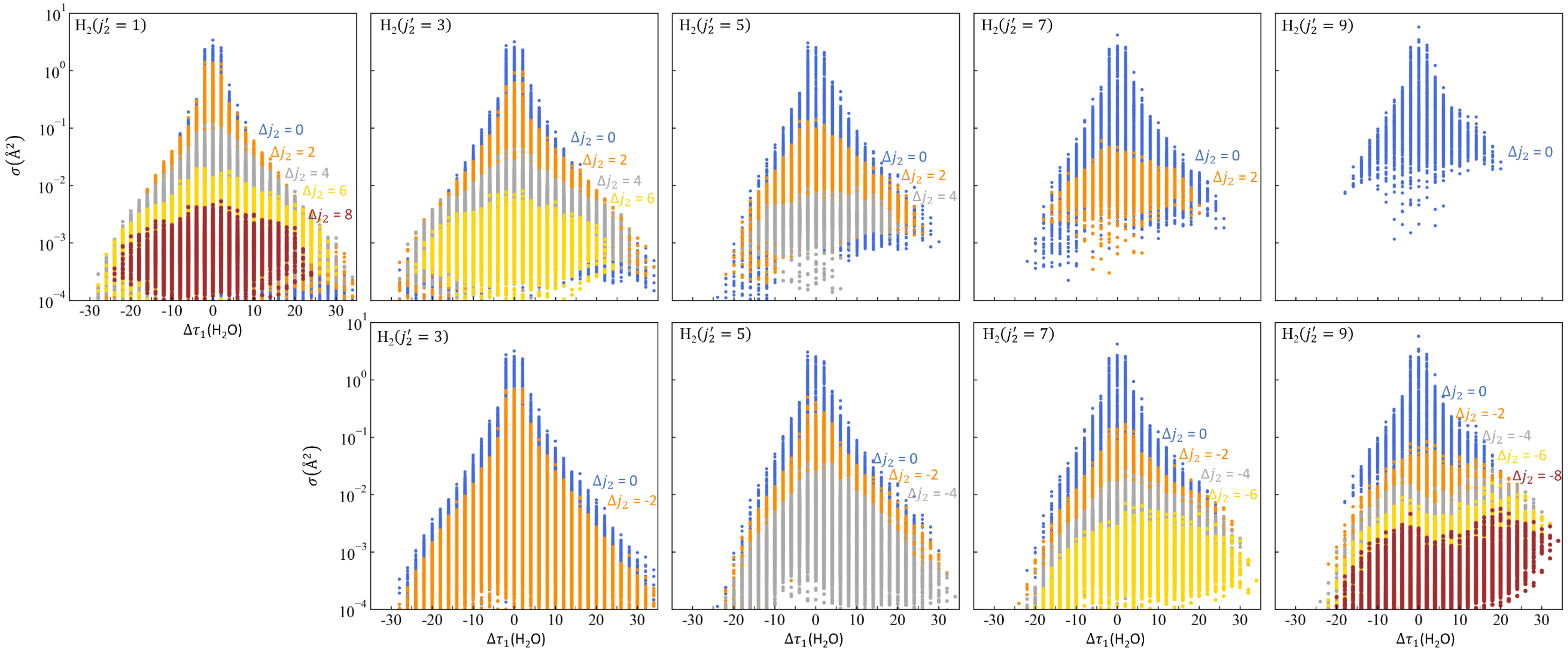

**Figure S19:** Correlation between the values of individual state-to-state transition cross sections $\sigma_{n'\to n''}$ and change in the $\tau_1$ of $H_2O$ ($\Delta\tau_1$) in the *p*-$H_2O$ + *o*-$H_2$ system, at $U$ = 12000 cm$^{-1}$. Here 100 initial states of water are considered. The initial state of $H_2$ molecule is given in the upper left corner of each frame. Top and bottom rows of frames correspond to excitation and quenching of $H_2$. Colors represent different transitions in $H_2$: $\Delta j_2 = 0, \pm 2, \pm 4, \pm 6, \pm 8$ are shown by blue, orange, grey, yellow, and maroon symbols, respectively.

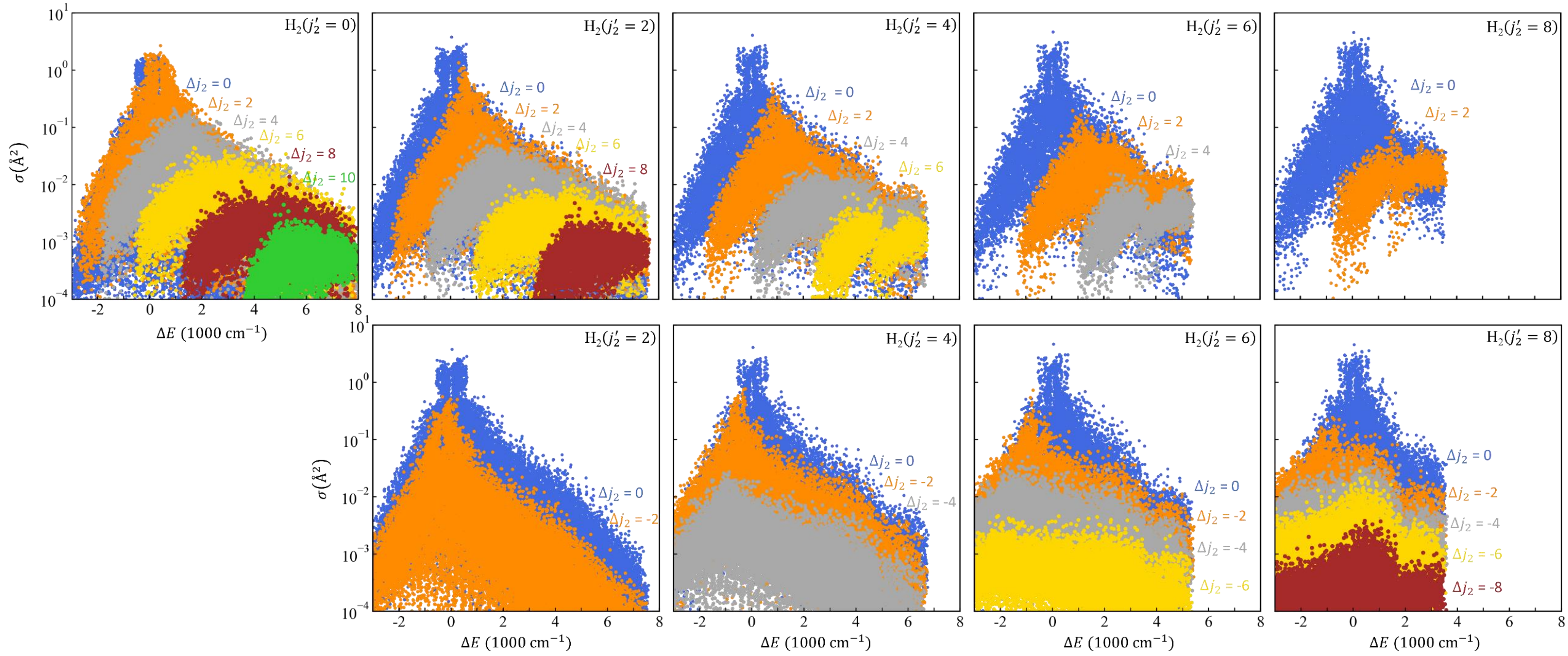

**Figure S20:** Correlation between the values of individual state-to-state transition cross sections $\sigma_{n'\to n''}$ and the overall transfer of internal rotational energy, $\Delta E$ in the *o*-$H_2O$ + *p*-$H_2$ system at collision energy $U$ ~ 12000 $cm^{-1}$. Here 100 initial states of water are considered. The initial state of $H_2$ molecule is given in the upper right corner of each frame. Top and bottom rows of frames correspond to excitation and quenching of $H_2$. Colors represent different transitions in $H_2$: $\Delta j_2 = 0, \pm 2, \pm 4, \pm 6, \pm 8$ are shown by blue, orange, grey, yellow, and maroon symbols, respectively.

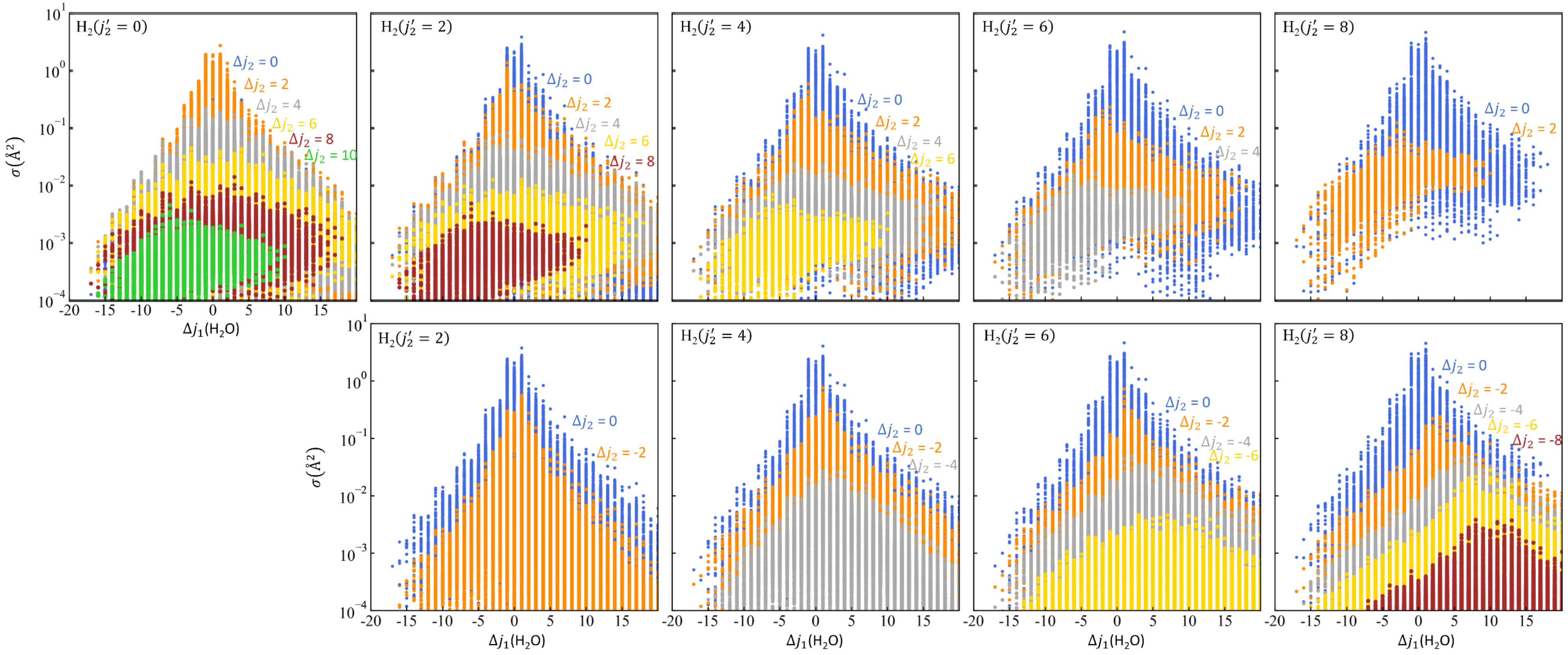

**Figure S21:** Correlation between the values of individual state-to-state transition cross sections $\sigma_{n' \to n''}$ and change in the rotational state of $H_2O$ ($\Delta j_1$) in the *o*-$H_2O$ + *p*-$H_2$ system at collision energy $U \sim 12000$ cm$^{-1}$. Here 100 initial states of water are considered. The initial state of $H_2$ molecule is given in the upper right corner of each frame. Top and bottom rows of frames correspond to excitation and quenching of $H_2$. Colors represent different transitions in $H_2$: $\Delta j_2 = 0, \pm 2, \pm 4, \pm 6, \pm 8$ are shown by blue, orange, grey, yellow, and maroon symbols, respectively.

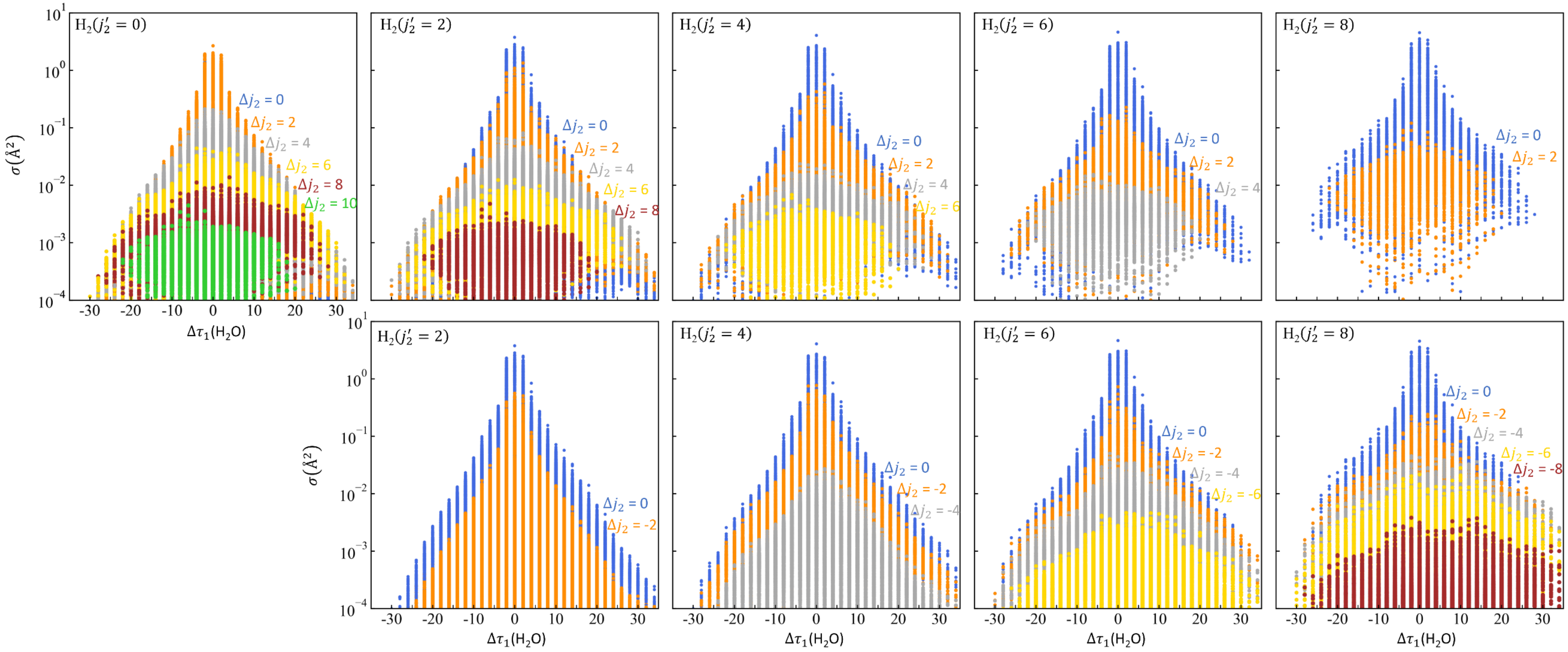

**Figure S22:** Correlation between the values of individual state-to-state transition cross sections $\sigma_{n'\to n''}$ and change in the $\tau_1$ of $H_2O$ ($\Delta\tau_1$) in the *o*-$H_2O$ + *p*-$H_2$ system at collision energy $U$ ~ 12000 cm$^{-1}$. Here 100 initial states of water are considered. The initial state of $H_2$ molecule is given in the upper left corner of each frame. Top and bottom rows of frames correspond to excitation and quenching of $H_2$. Colors represent different transitions in $H_2$: $\Delta j_2 = 0, \pm 2, \pm 4, \pm 6, \pm 8$ are shown by blue, orange, grey, yellow, and maroon symbols, respectively.

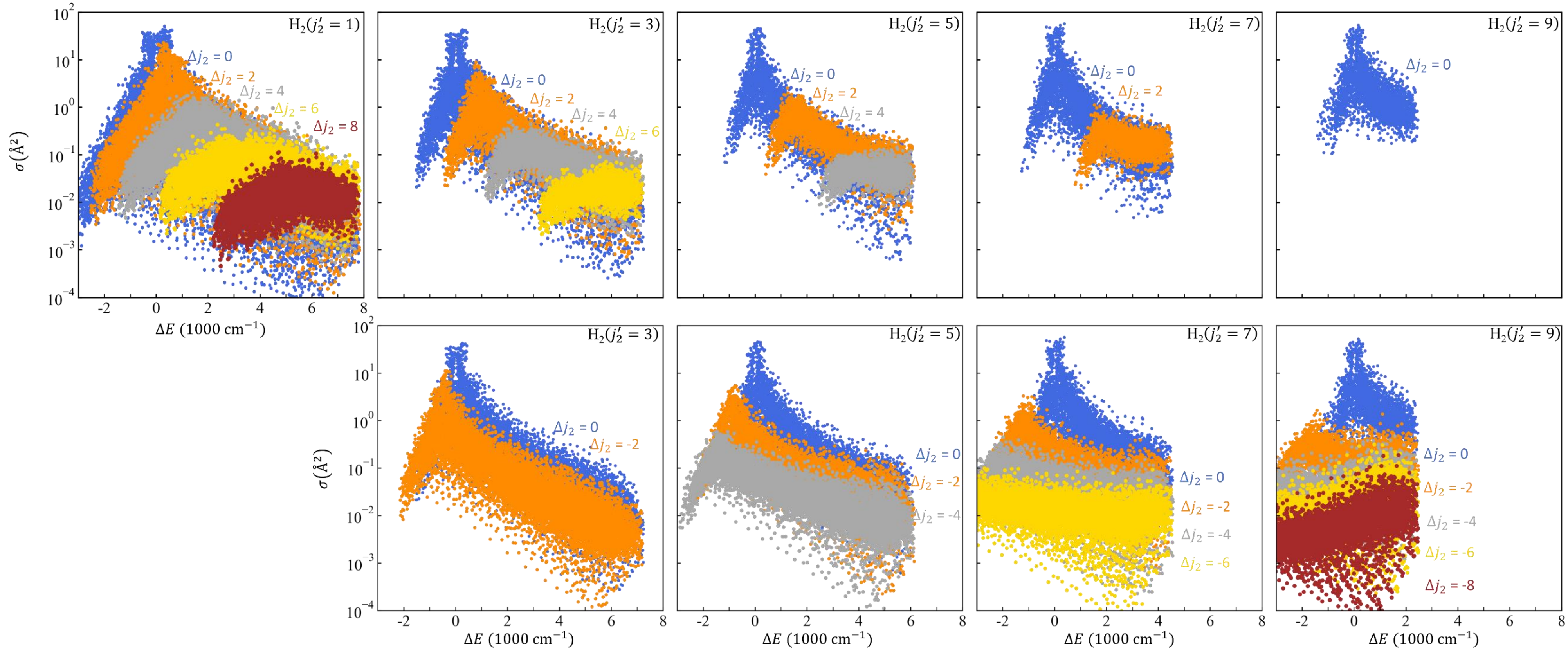

**Figure S23:** Correlation between the values of individual state-to-state transition cross sections $\sigma_{n' \to n''}$ and the overall transfer of internal rotational energy, $\Delta E$ in the *o*-$H_2O$ + *o*-$H_2$ system at collision energy $U \sim 12000$ cm$^{-1}$. Here 100 initial states of water are considered. The initial state of $H_2$ molecule is given in the upper right corner of each frame. Top and bottom rows of frames correspond to excitation and quenching of $H_2$. Colors represent different transitions in $H_2$: $\Delta j_2 = 0, \pm 2, \pm 4, \pm 6, \pm 8$ are shown by blue, orange, grey, yellow, and maroon symbols, respectively.

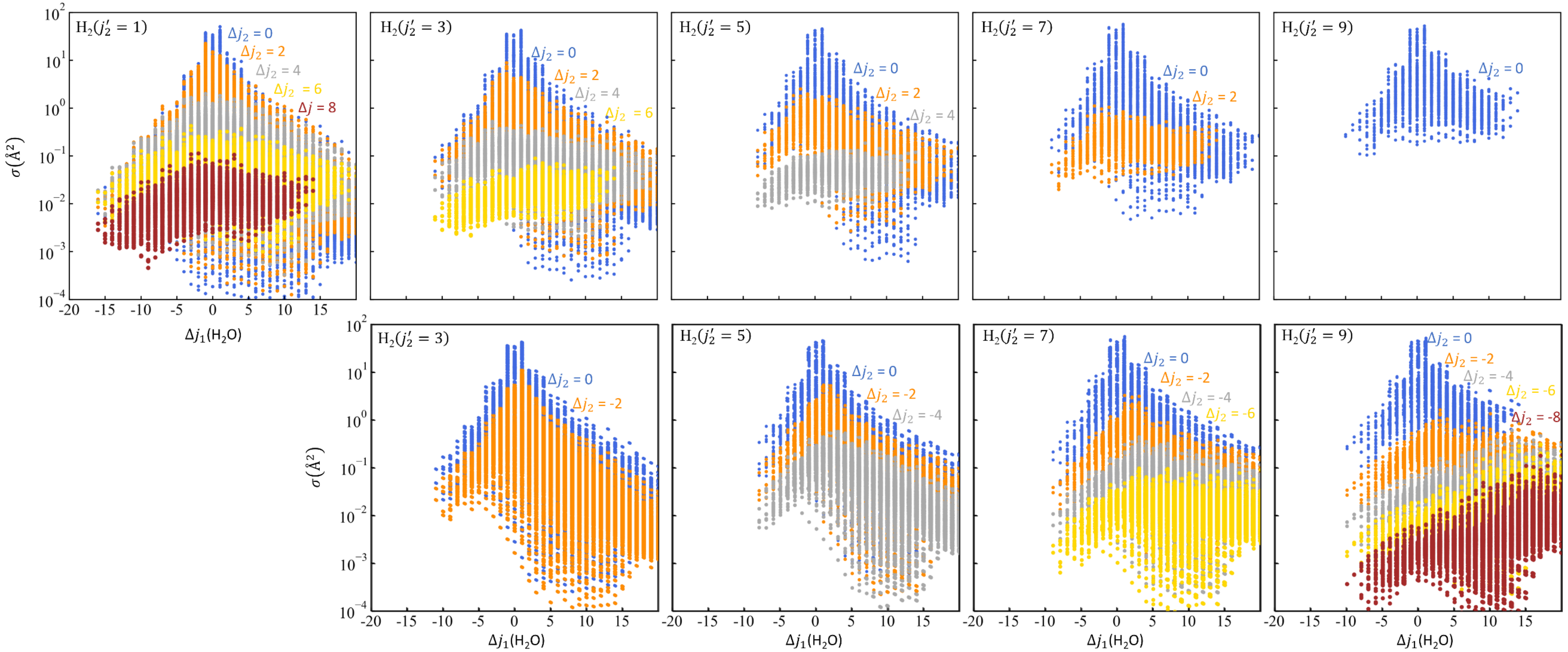

**Figure S24:** Correlation between the values of individual state-to-state transition cross sections $\sigma_{n' \to n''}$ and change in the rotational state of $H_2O$ ($\Delta j_1$) in the $o$-$H_2O$ + $o$-$H_2$ system at collision energy $U$ ~ 12000 $cm^{-1}$. Here 100 initial states of water are considered. The initial state of $H_2$ molecule is given in the upper left corner of each frame. Top and bottom rows of frames correspond to excitation and quenching of $H_2$. Colors represent different transitions in $H_2$: $\Delta j_2 = 0, \pm 2, \pm 4, \pm 6, \pm 8$ are shown by blue, orange, grey, yellow, and maroon symbols, respectively.

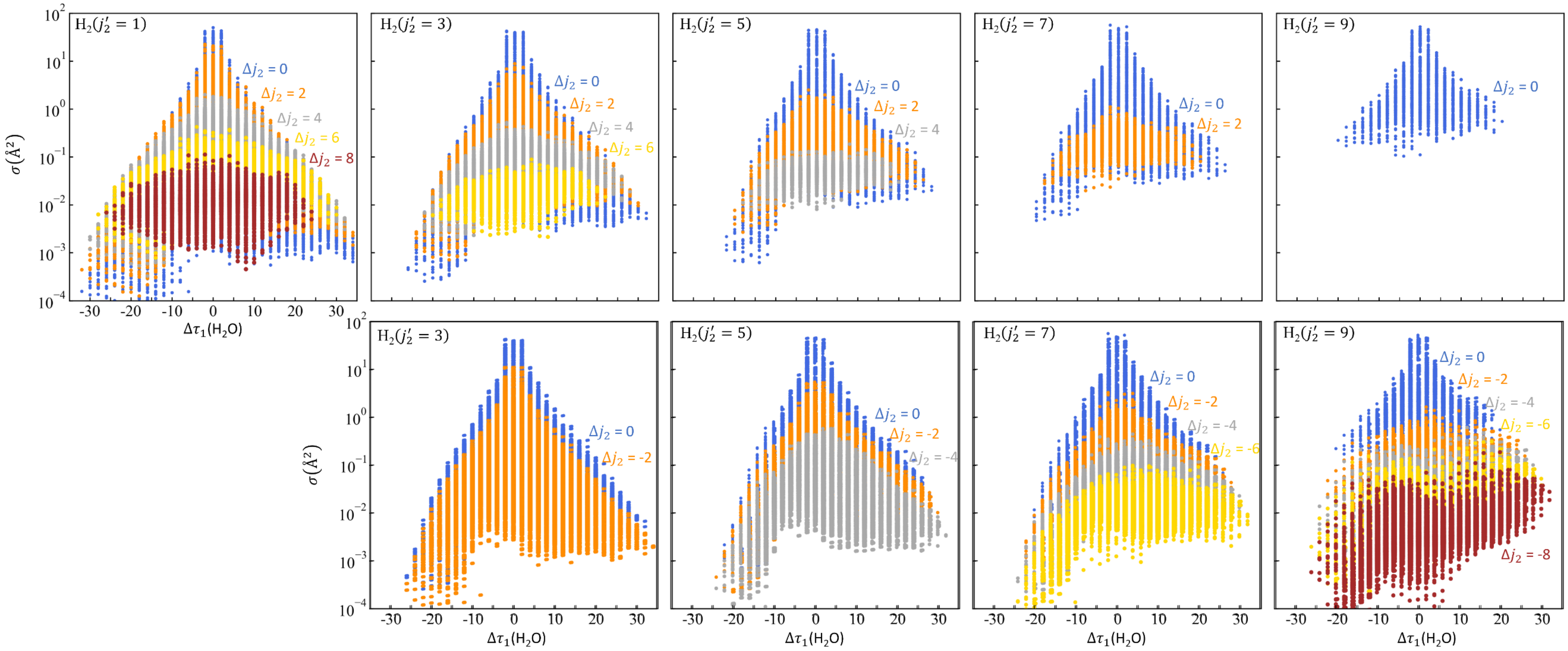

**Figure S25:** Correlation between the values of individual state-to-state cross sections $\sigma_{n'\to n''}$ and change in the $\tau_1$ of $H_2O$ ($\Delta\tau_1$) in the *o*-$H_2O$ + *o*-$H_2$ system at collision energy $U$ ~ 12000 cm$^{-1}$. Here 100 initial states of water are considered. The initial state of $H_2$ molecule is given in the upper left corner of each frame. Top and bottom rows of frames correspond to excitation and quenching of $H_2$. Colors represent different transitions in $H_2$: $\Delta j_2 = 0, \pm 2, \pm 4, \pm 6, \pm 8$ are shown by blue, orange, grey, yellow, and maroon symbols, respectively.